%% file: main.tex
\documentclass[authordraft]{acmart}

\AtBeginDocument{%
  }

\usepackage{float}
\usepackage{appendix}
\usepackage{xcolor} 
\usepackage[colorinlistoftodos, textsize=tiny, backgroundcolor=yellow!20]{todonotes}
\usepackage{listings}
\newcommand{\ConditionA}{\emph{Condition A (Provocations)}}
\newcommand{\ConditionB}{\emph{Condition B (No Provocations)}}\em
\begin{document}

\title{``It makes you think'': Provocations Help Restore Critical Thinking to AI-Assisted Knowledge Work}

\author{Ian Drosos}
\email{t-iandrosos@microsoft.com}
\affiliation{%
  \institution{Microsoft Research}
  \city{Cambridge}
  \country{UK}
}

\author{Advait Sarkar}
\email{advait@microsoft.com}
\affiliation{%
  \institution{Microsoft Research}
  \city{Cambridge}
  \country{UK}
}
\affiliation{%
  \institution{University of Cambridge}
  \city{Cambridge}
  \country{UK}
}
\affiliation{%
  \institution{University College London}
  \city{London}
  \country{UK}
}

\author{Xiaotong (Tone) Xu}
\email{xt@ucsd.edu}
\affiliation{%
  \institution{University of California, San Diego}
  \city{La Jolla}
  \state{California}
  \country{USA}
}

\author{Neil Toronto}
\email{neil.toronto@microsoft.com}
\affiliation{%
  \institution{Microsoft Research}
  \city{Cambridge}
  \country{UK}
}


\begin{abstract}
Recent research suggests that the use of Generative AI tools may result in diminished critical thinking during knowledge work. We study the effect on knowledge work of provocations: brief textual prompts that offer critiques for and propose alternatives to AI suggestions. We conduct a between-subjects study ($n=24$) in which participants completed AI-assisted shortlisting tasks with and without provocations. We find that provocations can induce critical and metacognitive thinking. We derive five dimensions that impact the user experience of provocations: task urgency, task importance, user expertise, provocation actionability, and user responsibility. We connect our findings to related work on design frictions, microboundaries, and distributed cognition. We draw design implications for critical thinking interventions in AI-assisted knowledge work.
\end{abstract}



\begin{teaserfigure}
  \includegraphics[width=\textwidth]{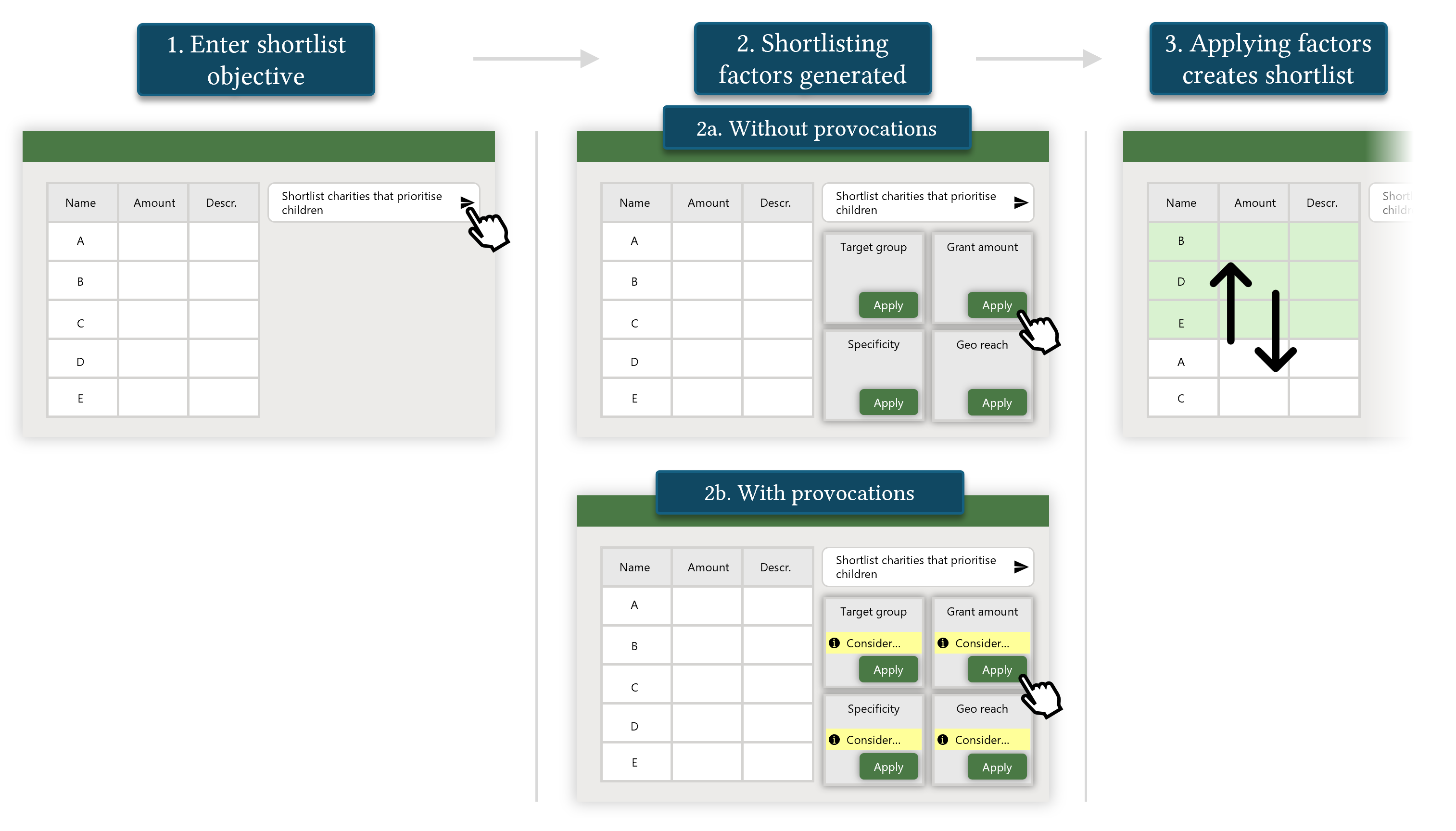}
  \caption{Schematic overview of our shortlisting interface for evaluating the impact of AI-generated provocations on critical thinking. 1. The user loads their dataset and enters an overall objective for the shortlist. 2. The system generates a set of candidate factors that can be applied to rank rows. 2a. In this experimental condition, no provocations are shown. 2b. In this experimental condition, factors are shown with ``provocations'' (indicated in yellow), system-generated critiques that highlight limitations, risks, and biases of the suggested factor. 3. Applying factors causes the system to re-order the dataset and identify a shortlist (indicated in light green) based on all currently applied factors.}
  \Description{A series of labelled mockups of a spreadsheet-like interface showing a data grid, text box where users enter their shortlist objective, smaller boxes for displaying shortlisting factors, and an illustration representation of grid rows being sorted.}
  \label{fig:overview}
\end{teaserfigure}


\maketitle

\section{Introduction}
\label{sec:Introduction}
Generative AI (GenAI), defined as any \emph{``end-user tool [...] whose technical implementation includes a generative model based on deep learning''}\footnote{This is a nebulous term with many potential interpretations and connotations. For clarity we adopt this definition, a rationale for which is given by \cite{sarkar2023eupgenai}.}, is achieving wide penetration in knowledge workflows. As a result, knowledge work is shifting \emph{``from material production to critical integration''} \cite{sarkar2023aiknowledgework}. This means that the physical activities of producing text, images, video, audio, etc. such as typing and drawing are being increasingly delegated by users to GenAI tools, and the user focus shifts to deciding when, where and how to apply GenAI to their work, and evaluating and editing its output to integrate it back into the ecosystem of their knowledge artefacts \cite{tankelevitch2024metacognitive, bodker2012dynamics, singh2023hide}.

But this shift comes at a cost to human critical thinking (defined in Section~\ref{sec:related-work-CT}). One obvious risk is \emph{overreliance}, defined as \emph{``as users accepting incorrect AI recommendations -- i.e., making errors of commission.''} \cite{passi2022overreliance}. But beyond errors and hallucinations, there is a subtler and more pervasive tendency that affects knowledge work even when the output is ``correct'' in some objective sense. Recent research has identified a tendency toward ``mechanised convergence'' \cite{sarkar2023aiknowledgework,sarkar2024intention}, that the introduction of GenAI to a knowledge workflow increases the homogeneity of work produced. There is substantial evidence across multiple studies and knowledge workflows for the phenomenon of mechanised convergence. Predictive text encourages predictable writing \cite{arakawa2024coaching-copilot}. Programmers using GitHub Copilot produce fewer unique identifier names \cite{lee2024predictability}. Exposure to GenAI ``ideas'' leads to less diverse content produced by short story writers \cite{Doshi2023GenerativeAI}. Strategy consultants using ChatGPT retain a high proportion of its response, and produce less conceptually varied ideas than those without access \cite{dell2023navigating}. Participants produce less distinct ideas in a creative ideation task when using ChatGPT than without it \cite{anderson2024homogenization}. Visual artists using GenAI experience a decline in average visual and content novelty \cite{zhou2024generative}. Conversational search can produce a ``generative echo chamber'' \cite{sharma2024echo}. Co-writing with ``opinionated'' language models shifts writers' opinions \cite{jakesch2023opinionated}.


People tend to offload effortful cognitive reasoning tasks to technology \cite{barr2015brain}. Deliberate thinking requires mental resources that people are unwilling to devote \cite{westgate2017little}, and media with fewer opportunities for interaction contribute less to the development of critical thinking \cite{saade2012critical}. Analytic methods from economics calculate that a significant proportion of jobs have significant exposure to (i.e., possibility of automation with) GenAI \cite{eloundou2024gpts}, a trend borne out in surveys of GenAI adoption \cite{bick2024rapid}. The complementary risks of overreliance and mechanised convergence build toward a picture where GenAI has a negative effect on knowledge work by compromising the ability of human workers to form deeply personalised, critical, contextual, and subjective responses to AI output in the process of integrating it into their workflow.

In this paper, we evaluate the concept of \emph{provocations}: short text prompts whose aim is to induce critical thinking in the user in AI-assisted tasks. Provocations highlight risks, biases, limitations, and alternatives of GenAI suggestions. Provocations draw on two lineages of research: explorations of integrating critical thinking aids in education (detailed in Section~\ref{sec:related-work-CT}), and design research for promoting reflective thinking (Section~\ref{sec:related-work-design}). While much research has been conducted in improving critical thinking in education and certain professional disciplines, interventions for critical thinking in contemporary, AI-assisted knowledge workflows have not been studied. Similarly, design research in reflective thinking focuses on applications such as promoting mental health and wellbeing, misinformation detection, and media consumption habits, but critical thinking interventions for common knowledge workflows remain under-explored.

We focus on one such workflow: \emph{shortlisting} (detailed in Section~\ref{sec:shortlisting_motivation}). For example, shortlisting candidates to interview from a list of job applicants, choosing a set of houses to view from a list of search results, or shortlisting papers to read. Shortlisting is a common yet important and diverse type of knowledge work where GenAI can be applied to suggest factors for shortlisting and automatically evaluate candidate items according to these factors. Yet shortlisting encompasses many detailed, conscious, subjective and critical judgements, which makes it important to explore how AI can be used to support the task without diminishing -- and preferably, enhancing -- the user's ability to think critically about their problem.

We make the following contributions:

\begin{itemize}
    \item We provide a brief overview of the literature on critical thinking and situate our work within the space of design interventions for critical and reflective thinking (Section~\ref{sec:related-work}).
    \item We instantiate the idea of provocations in a prototype for shortlisting, a common knowledge workflow requiring critical and subjective judgment (Section~\ref{sec:SystemDesign}).
    \item We present a user study ($n=24$) comparing the user experience of AI-assisted shortlisting with and without provocations (Section~\ref{sec:Methods}). We show how provocations can counteract the tendency for mechanised convergence, and induce critical thinking at all levels of Bloom's taxonomy. We derive a set of salient dimensions of the user experience of provocations (Section~\ref{sec:Results}). Our primary contribution is in understanding the human experience of provocations, rather than contributing a tool or design guidelines.
    \item We discuss how our work complements, yet throws new light on previously identified design challenges in critical thinking interventions. We address difficulties in quantifying the impact of provocations, and propose implications for the design of future critical thinking tools (Section~\ref{sec:discussion}).
\end{itemize}

\section{Background and Related Work}
\label{sec:related-work}

\subsection{Critical Thinking}
\label{sec:related-work-CT}



\subsubsection*{Frameworks}

``Critical thinking'' is a contested term. As a starting point, we adopt the influential framework developed by Bloom et al. \cite{bloom1956taxonomy, huitt2011bloom}, a hierarchical taxonomy that characterises student learning objectives into six types: knowledge (recall of ideas), comprehension (demonstrating understanding of ideas), application (putting ideas into practice), analysis (contrasting and relating ideas), synthesis (combining ideas), and evaluation (judging ideas through criteria). 

Subsequent scholarship has developed similar frameworks and also identified psychological dispositions (not unlike personality traits) such as inquisitiveness, that mediate critical thinking \cite{facione1990critical, facione2011critical, facione1995disposition, paul1997california}. Though not all scholars use the terms interchangeably, critical thinking is often also referred to as reflective thinking \cite{dewey1910howwethink}. Research in reflective thinking often conceptualises it as a skill that is developed in sequential stages \cite{nguyen2014reflection,king1997reflective,paul1997california}. There have been multiple proposals to integrate these disparate frameworks for critical thinking \cite{kuhn1993connecting, mulnix2012thinking, dwyer2014integrated}. However, in our research we adopt the Bloom et al. framework for its relative simplicity, as well as its strong support in the research literature and wide operationalisation in educational systems.

Despite concerns about whether critical thinking can be taught \cite{willingham2008critical}, research in education has developed a number of approaches to teaching critical thinking. These involve exercises in argumentation and metacognitive thinking \cite{king1997reflective,wilson2016teaching,paul1997california}, or applying formal methods of argument analysis such as the Toulmin model of argumentation \cite{kneupper1978teaching} and argument mapping \cite{davies2011concept}, often operationalised in software tools \cite{tsai2015argumentation}.

Critical thinking can be measured, or assessed, either through self-evaluation or by the evaluation of peers or expert assessors. Researchers have developed several instruments, such as questionnaires, justified multiple choice questions, structured essays, protocols for whole-portfolio assessment, task observation, and peer interaction \cite{ennis1993critical,PAUL20141357,kobylarek2022critical,kember2008four, facione1994critical,zuriguel2017development,zuriguel2022nursing}. \citet{kember2000development} developed a widely deployed self-assessment questionnaire, which we adopt in our work.

%
%

An important influence on the design of our tool comes from Salomon's work in computer-assisted learning \cite{salomon1988ai}, which showed that periodically posing critical questions such as \emph{``what kind of image have I created from the text?''} resulted in an enduring improvement in students' reading comprehension. We extend the idea of the critical question into our provocations, by applying the questioning to AI output, and further, using a large language model to generate critiques that are more task-specific. 




\subsection{Designing for Critical Thinking and Reflection}
\label{sec:related-work-design}

\subsubsection{Design Dimensions}
One design dimension for critical or reflective thinking interventions is whether the system should be proactive, i.e., introduce critical thinking prompts without an explicit user request. There is mixed evidence that suggests the appropriate solution is heavily context dependent. For example, in a stock investment workflow, the system might only intervene when the user is about to make an important decision \cite{reicherts2022chatbots}. On the other hand, in a language micro-learning workflow, the system might interrupt the user at random occasions \cite{kim2024microlearning}.



Research has demonstrated the importance of user participation in critical thinking interventions. For instance, \citet{danry2023ask} find that AI explanations presented as questions, rather than statements, improves logical discernment. Surfacing contextual questions and encouraging the self-generation of questions improves critical engagement with reading \cite{yuan2023critrainer, maldonaldo2023readerquizzer}. Attention checks during surveys promote systematic thinking on tricky tasks \cite{hauser2015instruction-checks}. Engagement with artistic artefacts and conflict-filled peer-discussion fosters critical thinking about AI in youth \cite{lee2023fostering}. Increased participation and engagement in the reasoning process, including with other human discussants, is more likely to result in attitudinal and behavioural changes \cite{miller2001counter, liao2023deepthinkingmap}.

Another design dimension is the extent to which critical thinking interventions are agentised or presented in an anthropomimetic manner. Different styles of anthropomimesis can significantly impact the user experience of interfaces for critical reflection, such as AI personas embodying distinct demographics \cite{yeo2024self-reflection}, or in-group vs. outgroup identity or persuasive vs. argumentative styles \cite{tanprasert2024debate-chatbots}, and can lead to a shared locus of control \cite{nordberg2023conversations-news}.


Many reflective thinking interventions are presented as games or include elements of gamification, suggesting a ludic element to reflection (e.g., \citet{tang2024mystery, du2024careersim}). \citet{miller2024reflectiveplay} synthesize a design framework for reflective play, identifying key elements from games that promote reflection, namely: disruptions, slowdowns, questioning, revisiting, and enhancers.

\subsubsection{Application Domains}
A common application domain for critical thinking applications is in preventing online misinformation and propaganda. Research has demonstrated the effectiveness of approaches such as structured thinking aids \cite{holzer2015mobile,holzer2018debate}, nudges that aim to activate the analytical mode of thinking in Kahneman's dual-system theory of cognition \cite{zavolokina2024propaganda}, worksheets and group discussion \cite{wang2024koala}, and gamification of misinformation detection \cite{tang2024mystery}.

Another setting where critical thinking interventions have been studied is in writing, ideation and argumentation tools. Applications for visualising and developing structured arguments improve critical thinking in individual and collaborative settings \cite{sun2017critical, tsai2015argumentation}, can help dislodge users from a ``self-imposed filter bubble'' \cite{aicher2023self-bubble}, and help writers of speculative fiction reflect on future scenarios \cite{tost2024futuring-machines}. LLM-driven writing support tools for the ideation and evaluation processes (of the Cognitive Process Theory of writing) can increase the duration of engagement with writing, but in-task writing support is used less often when static support is enabled by default \cite{goldi2024writers-cognitive}. LLM-generated reflective prompts can be inserted into ideation templates for digital whiteboards; the spatial layout inherent to whiteboarding allows the reification of relationships and hierarchy between pieces of information, an advantage over solely chat-based interaction \cite{xu2024jamplate}. Access to Generative AI can be beneficial in the divergent phase of ``brainwriting'' by surfacing ideas that are novel (to the writer) \cite{shaer2024brainwriting}. Access to a chatbot interface for assessing risks can help researchers write impact statements more critically, but can be viewed as a single source of guidance and create a false sense of due diligence \cite{mukherjee2023impactbot}. Shibani et al. propose a framework for assessing student writers' critical interactions with AI assistants, offering a rubric for assessment along dimensions such as planning and ideation, information seeking and evaluation, writing and presentation, personal reflection, and conversational engagement \cite{shibani2024critical-learner}.

Yet another common application is in mental health and wellbeing. Kitson et al. synthesise a framework of technological interventions for cognitive reappraisal, a process of active reflection on a situation to decrease its emotional impact and improve mental health \cite{kitson2024cognitive-reappraisal}. Reflective prompts can help reduce compulsive smartphone use \cite{li2023stayfocused-formative, li2024stayfocused} and improve time management \cite{jeromela2023time-management}. However, misalignment between perceived habits and behavioural logs (e.g., desire to read serious, in-depth news misaligns with recorded behaviour of reading less serious news) can lead to negative reflections \cite{aubinlequere2024news}. LLMs can be applied to narrativise time series behavioural patterns (e.g., sleep, activity, location) and create journaling prompts \cite{nepal2024mindscape}, or encourage reflection on e-book highlights \cite{kang2024quologue}, or support spirituality and prayer \cite{kwon2024spiritual}. They can generate ideas and resources to support self-care \cite{capel2024self-care}, and facilitate coach-client relations in executive coaching by supporting self-reflection for leadership growth \cite{arakawa2024coaching-copilot}. LLM-generated questions based on narratives stored in NFC-tagged objects can support active recollection and reflection on cherished objects \cite{jeung2024treasurefinder}. Physical externalisation of behavioural data, such as creating printed ``attention receipts'' for time spent engaging with YouTube videos, can help with reflection \cite{sathya2024attention-receipts}. However, when tools lack appropriate context, they can lead to reflections that ``felt akin to horoscope readings with vague and impersonal responses'' \cite{yo2024autoethnographic}. 

Our application domain of shortlisting with datasets (Section~\ref{sec:shortlisting_motivation}) is a particular type of data analysis, or sensemaking activity \cite{pirolli2005sensemaking,drosos2024rubberduck}. Critical thinking support for data analysis remains relatively under-explored, particularly when the analysis is AI-assisted. Previous work has explored integrating notifications into computational notebooks to highlight fairness and bias issues for data scientists \cite{harrison2024jupyterlab-notifications}. A study of GitHub Copilot use reveals that while users often accept suggestions without verifying them, this is done to take advantage of certain IDE features (such as syntax highlighting) and suggestions are subsequently verified; yet it is unclear if verification merely entails checking for correctness, rather than quality (the latter would require more critical thinking) \cite{mozannar2024reading}. Considering image retrieval in art history as a form of data-driven sensemaking, Glinka and M\"{u}ller-Birn propose four implications for ``critical-reflective human-AI [interaction]'': supporting reflection on the basis of transparency, foregrounding epistemic presumptions, emphasizing the situatedness of data, and strengthening interpretability through contextualized explanations \cite{glinka2023art-history}.

\textbf{In summary}, previous work has explored how to define and teach critical thinking. Moreover, design research has explored critical thinking interventions in a variety of domains, such as online misinformation, writing, ideation, mental health, and data analysis. However, research has not yet empirically evaluated any interventions that aim to counteract the mechanised convergence tendencies of GenAI-assisted workflows through critical thinking aids. This is the gap we address. Our primary aim is to understand the human experience of critical thinking aids in AI-assisted workflows, rather than contributing a tool or design guidelines.

\section{System Design}
\label{sec:SystemDesign}

\subsection{Shortlisting}
\label{sec:shortlisting_motivation}
While our work is motivated by the tendency towards overreliance and mechanised convergence in knowledge work broadly, to focus our investigation it is necessary to narrow our scope to specific tasks. We chose \emph{shortlisting} -- selecting a small set of items from a larger set -- as the user task on which to focus our investigation. Shortlisting tasks are a ubiquitous type of knowledge work: examples include shortlisting interview candidates from a list of job applicants, shortlisting papers to accept for a conference from a list of submissions, shortlisting software features to implement from a list of user requests, etc. 

As a type of sensemaking through data analysis \cite{pirolli2005sensemaking}, shortlisting is relatively small in scope: it involves choosing relevant factors and their weights for evaluating candidates, then applying those factors to individual candidates, sorting and ranking them. While small in scope, shortlisting activities are nonetheless diverse and ecologically valid \cite{sarkar2024copilot}. They present a suitable test bed for evaluating provocations because they require users to enact a range of critical thinking behaviours: choosing qualitative and quantitative factors, applying them (often subjectively) to candidates, and iterative decision-making and reflection as users refine their factors and shortlist selection over the course of the task.

Shortlisting is not the only such task; several types of data-driven sensemaking tasks are diverse yet well-constrained, such as personal budgeting and financial planning, online shopping and comparison, etc. \cite{drosos2024rubberduck}. Suitable tasks are also available in other domains, such as design or writing. As a research team, our collective prior experience in qualitative analysis has been focused on studies of end-users performing data analysis tasks in spreadsheets, including shortlisting, and our relative strength in qualitative interpretation of participant experiences in such tasks informed our choice. 


\subsection{Design Decisions and Research Goals}
Our system takes two roles during shortlisting. The first is the conventional role of AI as \emph{assistant}. The system assists by suggesting potential factors and weights for those factors, that the user may wish to use to evaluate the candidate items in their dataset. Moreover, the system evaluates each item according to the factors selected by the user, sorts the list of items, and creates a shortlist of the highest ranking items. This process is illustrated in Figure~\ref{fig:overview}.

The second role is that of AI as \emph{provocateur} \cite{sarkar2024challenge} or critical thinking tool \cite{ye2024languagemodelscriticalthinking} -- presenting provocations that aim to induce critical thinking about the use of AI suggestions in the user's workflow. This is the role that we are interested in evaluating.

The design of provocations in our system can be seen in Figure~\ref{fig:factor_cards}, and their role within the overall shortlisting workflow in Figure~\ref{fig:overview}. Our design is informed by the following principles from prior literature:

\begin{itemize}
    \item From \citet{salomon1988ai} we apply the key technique of presenting short texts to prompt user reflection. We limit provocations to two sentences to avoid information overload, according to guidelines and best practices for human-AI interfaces \cite{amershi2019guidelines}.
    \item Due to the complexity and range of phenomena associated with agentised or anthropomimetic presentation (e.g., \cite{yeo2024self-reflection,tanprasert2024debate-chatbots,nordberg2023conversations-news}) we present provocations in the language of plain factual warnings, to avoid any inadvertent confounding effects of agentic presentation.
    \item Previous research on the timing and quantity of interventions has shown that the solution depends on the context \cite{reicherts2022chatbots,kim2024microlearning}; we attach a single provocation directly to each AI-generated factor suggestion, presented at the same time and location as the suggested factor. This allows provocations to apply to individual suggestions and thus be more specific, while only slightly increasing the cognitive effort of consuming the information related to each factor; the concurrent presentation allows the user to encounter and consider suggestion-provocation pairs in any order and at any time. 
\end{itemize}

Clearly, many aspects of our design could be varied, such as provocation tone, grammar, different prompting strategies for generating texts, presenting provocations as diagrams, images, or video, making provocations interactive, etc. Even the choices we have made above, such as avoiding agentic presentation, could be varied. The implicit research question behind such variations is: \emph{how can we make provocations more effective through design?}

However, we are not (yet) interested in evaluating very specific questions about the design of provocations. We are exploring a question that is a precursor, namely: whether provocations -- known to be effective for improving critical thinking in an educational setting, targeting students, without AI assistance -- can also be effective for professionals conducting knowledge work with AI assistance. And if they are, what are the mechanisms by which they act, and what aspects of the workflow most impact their effectiveness? These questions are restated as more concise research questions in Section~\ref{sec:Methods}. If provocations on the whole turn out to be ineffective for some unanticipated reason that is uncovered through empirical study, then it will have been neither interesting nor important to investigate minor variations in visual presentation. In the terms of \citet{tohidi2006getting}, we are at the stage of ``getting the right design'', which is anterior to ``getting the design right''.

\subsection{Implementation}

\begin{figure}
    \centering
  \includegraphics[width=\textwidth]{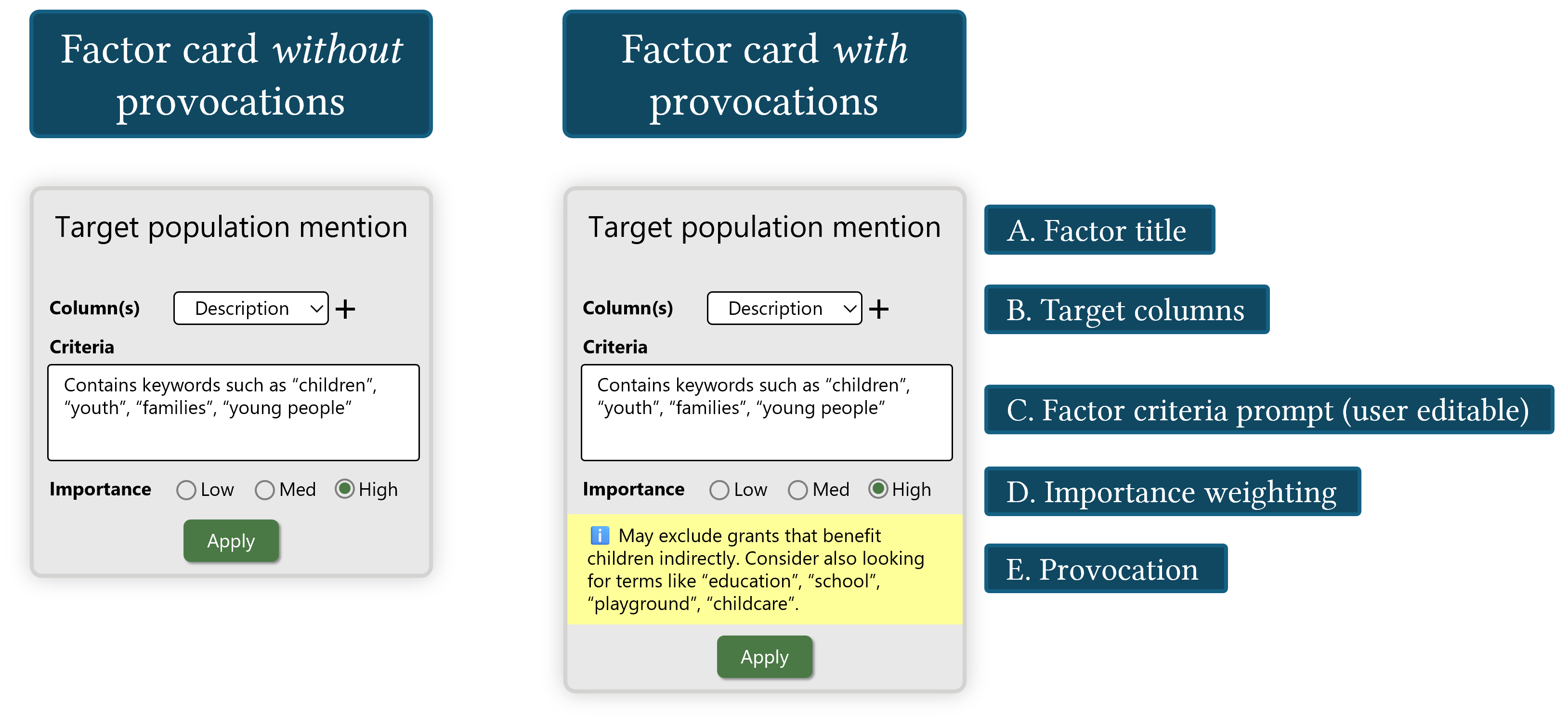}
  \caption{Factor cards without and with provocations.}
  \Description{}
  \label{fig:factor_cards}
\end{figure}

Users begin by loading their dataset into a spreadsheet-like interface, and entering an objective for their shortlist (Figure~\ref{fig:overview}-1). Based on the dataset and objective, the system generates a set of proposed factors displayed in factor cards (Figure~\ref{fig:factor_cards}), using a chain of calls to a large language model (details of the prompts used are given in Appendix~\ref{Appendix:Prompts:GenerateFactors}, \ref{Appendix:Prompts:GetImportance}). Users also have access to UI controls for deleting and manually adding factors.

A factor represents some desirable property, aiming to capture some aspect of the user's original intent, with which rows in the dataset may be evaluated for their suitability of inclusion in the shortlist. Our system generates a variety of factor types with a variety of attendant evaluation criteria, such as the presence of keywords in a text column, higher or lower values in a numerical column, and qualitative judgments of text fields (e.g., ``specificity of proposal''). The full list of pre-generated factors used in our study is given in Appendix~\ref{Appendix:FactorsGenerated}. For simplicity of implementation, our system applies LLM calls to generate all types of factors as well as perform all of these types of evaluations. We acknowledge that LLMs are inherently better or worse suited for some types of tasks and a system ``in production'' may approach this differently (e.g., for evaluating higher or lower numerical values, we might synthesize a short code snippet which would evaluate this more reliably and deterministically, rather than apply an LLM directly).

Each factor card contains the following UI elements:
    \begin{itemize}
        \item Title: A natural language description of the factor.
        \item Target columns: An editable list of columns from the dataset are leveraged to apply the factor criteria to the shortlist. Users can add or remove columns by interacting with the factor card.
        \item Factor criteria: The user-editable criteria that defines how the factor will be applied to the shortlist. This could be specific criteria (e.g., ``Amount applied for $<$ 10,000'') or open-ended instructions for the LLM (e.g., ``Charity asks for an amount that seems reasonable for what services they provide'').
        \item Factor provocation: The AI-generated provocation appears as text highlighted in a yellow box. Provocations are critiques of the factor. Provocations may include limitations, alternative factors, suggestions for criteria and importance, or may be open-ended without a clear way to apply the provocation.
        \item Factor importance: An editable rating from ``Low'' to ``High'' that represents how important the factor is to the shortlist decision. This assigns each factor $f$ an importance weight $w_f$ (High importance: $w_f = 1$, Medium: $w_f = 0.66$, Low: $w_f = 0.33$).
        \item Apply button: A button that applies the factor criteria to the shortlist. 
\end{itemize}


When the Apply button is pressed, the system uses an LLM call (prompts given in Appendix~\ref{Appendix:Prompts}) to evaluate whether each row satisfies the criteria as specified. The weighted factor score for each row is the sum of $w_f$ for all factors $f$ where the row satisfies $f$'s criteria. This score is used to rank rows and create a shortlist. This method prioritised implementation simplicity, but involves multiple ad-hoc, essentially arbitrary decisions. If the focus was on improving shortlisting \emph{per se}, this method should be determined more carefully. However, our focus is on the effect of provocations on critical thinking -- we leave detailed investigation of ranking methods out of scope. Future work may investigate other ranking and shortlisting methods that have been previously proposed in the contexts of human resources management, information retrieval, and recommender systems (e.g., \cite{bartell1994automatic,liu2014automatic,schnabel2016using, aydin2023ai}).



Within each shortlisted row, factors' target columns are coloured green to indicate that the system evaluated the column as satisfying the factor criteria.

\subsection{Example Usage Scenario}
\label{sec:usage-scenario}

For an example of how a user might interact with our system, consider Clara. Clara works for her local government's charity grant selection board, responsible for distributing grant money to charity organisations. This year's grant applications are stored in a spreadsheet, including data like the name of the charity, the amount of money being applied for, who the grant might help, and the description of what the charity will do with the money if funds are granted.

Clara's government has decided to prioritise funding grants focused on helping children. Clara uploads the dataset of grant applications to the system. She enters her shortlist objective into a text box by typing "Prioritise children" (Figure~\ref{fig:overview}, (1)). The system generates three factors for Clara to consider: "Target population", "Amount applied for", and "Specificity of the proposal".


Clara inspects the "Target population" card (Figure~\ref{fig:factor_cards}) and sees that it looks for keywords relating to children. She agrees with this but also sees an AI-generated provocation critiquing the factor, suggesting additional keywords like "school". Clara agrees and modifies the criteria to include these keywords.

Clara presses "Apply" to test how her changes affect the shortlist. Clara inspects the shortlist and moves on to the "Amount applied for" card. She agrees with the provocation that higher amounts might not mean greater impact and lowers the factor's importance to "Low" (Figure~\ref{fig:factor_cards}, (D)).


Clara continues with the "Specificity of the proposal" factor and reads a provocation suggesting that "more detailed proposals do not guarantee success". This prompts her to ideate a new factor called "Previous performance". However, her dataset lacks performance information. Realising that further data collection is needed, she plans to contact the shortlisted charities and collect information about their previous performance, and notes this reflection for future shortlists.

Satisfied with the results, Clara exports the shortlist for presentation to the grant selection board.

\section{Study Design}
\label{sec:Methods}
As motivated in Section~\ref{sec:Introduction}, the phenomena of overreliance and mechanised convergence can be interpreted as evidence of the tendency for knowledge workers to think less critically when using GenAI assistance. Informed by evidence from research in education and critical thinking interventions (Section~\ref{sec:related-work}), we implemented a prototype that instantiates the idea of \emph{provocations} in the context of GenAI-assisted shortlisting (Section~\ref{sec:SystemDesign}). We were interested in evaluating the effect of provocations in practice. In particular, our study addresses the following broad research questions:
\begin{itemize}
    \item[RQ1] Do provocations affect users' critical thinking during AI-assisted shortlisting tasks?
    \item[RQ2] What important dimensions of the system design, task, or user interact with the effectiveness of provocations during AI-assisted shortlisting tasks?
\end{itemize}

\subsection{Study Conditions}
\label{sec:Methods:Conditions}

Our study had two conditions: \ConditionA{} and \ConditionB{}. The only difference between these conditions is the presence or absence of provocation text in the factor cards (Figure~\ref{fig:factor_cards}).


We selected a between-subjects design for this study, assigning participants to two equal groups stratified by their GenAI and spreadsheet experience. We did not choose a within-subjects design because of the potential for strong priming effects from the provocations condition. In particular, since our hypothesis was that exposure to provocations induces critical thinking, it was possible that a participant exposed first to \ConditionA{} would carry over a critical disposition to tasks conducted in \ConditionB{}. We could have fixed the order so that \ConditionB{} always occurred first, but this would have resulted in order and learning effects.

Given our emphasis on qualitative analysis, and the fact that our core contribution is in understanding the effect of provocations on human thinking (as opposed to the specific interaction design of provocations), one alternative would have been to not frame this as a comparison at all, but rather have all participants use the system with provocations. It is reasonable to expect that provocations will induce critical thinking (given their established use for the same purpose in education), and thus it might have been more useful to study a larger sample of participants interacting with provocations.

Acknowledging this, we structured our study as a comparison to a without-provocations condition for two reasons. Firstly, the needs, motivations, priorities, and constraints of knowledge workers are different from those of students \cite{sarkar2024challenge}. Discretionary software users have a low tolerance for learning requirements or cognitive load, as identified by the ``paradox of the active user'' \cite{carroll1987paradox, carroll2014creating, sarkar2023simplicity}. The additional reading and thinking requirements of provocations may challenge participants' priorities for efficient task completion and desire or expectation for lower involvement (and indeed, we document this in Section~\ref{sec:Results:Dimensions}), meaning that there is value in empirical evidence that provocations do (or do not) induce critical thinking in knowledge work, compared to the same workflow without provocations. Secondly, a comparison to a without-provocations condition allows us to distinguish phenomena that arise as a specific consequence of provocations, given that think-aloud data can be intermittent and incoherent, and self-reflection is subject to known biases and potential for errors. A participant P may attribute some phenomenon (e.g., perception of cognitive load, or distrust, or confidence, or agency) to provocations, but if participants in the without-provocations condition also report the same phenomenon, this indicates that careful analysis is required to determine whether P's experience really was specific to provocations (and if so, how it differed from the without-provocations condition), or whether it was a misattribution.


\subsection{Tasks}
\label{sec:Methods:Tasks}
We designed two shortlisting tasks using two datasets, and a third task used only as a tutorial at the beginning to address learning effects.

For both tasks, participants were given the dataset and an initial shortlist objective, with the instruction to iterate through the generated factors to inspect, modify, and apply each card based on participant preference. 
 
Participants interacted with the tool until they were satisfied with their shortlist.

For consistency, all participants were given the same initial shortlist objective for both tasks (i.e., the text input in Figure~\ref{fig:overview}-1 was the same for all participants). This allowed us to pre-generate and cache the initial set of factor cards and provocations, so that each participant saw the same initial list of factors (and provocations for \ConditionA{}). However, the system was fully live and interactive for all subsequent interactions, and thus generated varying results based on participant edits to factors or addition of new factors.

\subsubsection{Task 1 (Charity board shortlist)}
Task 1 uses \emph{Dataset 1} which contains data about local grant applications of charities in the Cambridge, United Kingdom area. \emph{Dataset 1} contains columns for a unique identifier, the title of the grant which includes the charity's name, the description of the grant representing what the grant would be spent on, and the amount applied for.


The initial shortlist objective ``Prioritize children'' is used, which generates five factor cards (detailed in Appendix~\ref{Appendix:FactorsGenerated}): 
target population mention, amount of grant, specificity, geographical reach, and past performance.
Each factor card had corresponding criteria, recommended importance rating, and (for \ConditionA{}) provocation.

\subsubsection{Task 2 (Movie night shortlist)}
Task 2 uses \emph{Dataset 2} which contains data about movies on the Netflix streaming service. \emph{Dataset 2} contains columns for movie title, genre, tags, runtime, view rating, critic review ratings from IMDB, Rotten Tomatoes, and Hidden Gems, box office amount, summary, release date, director, and actors.


The initial shortlist objective ``Family movie night of bad movies'' is used which generates five factor cards (detailed in Appendix~\ref{Appendix:FactorsGenerated}): Target population mention, Amount of grant, Specificity of proposal, Geographical reach, and Past performance. Each factor card had corresponding criteria, recommended importance rating, and (for \ConditionA{}) provocation.

These tasks were designed to be ecologically valid, drawing on real shortlisting tasks encountered by the research team. They were designed to be complementary in two ways. First, the charity task represents a ``serious'' scenario with impactful consequences, whereas the movie night task is more ``casual''. Second, the charity task requires comprehension and domain reasoning skills in a relatively novel scenario, since participants had not encountered a charity shortlisting task prior to the study, whereas the movie night task allows participants to use their own prior knowledge in a familiar scenario. These complementary tasks thus represented two distinct levels of critical thinking needs and elicited a range of behaviours from participants.


\subsection{Participants}
\label{sec:Methods:Participants}

We recruited 24 participants (13 men, 11 women, 0 non-binary) via email from a list of spreadsheet users who had signalled interest in participating in user studies run by our institution. Our demographics form used previously developed questionnaire items and corresponding integer coding schemes for assessing experience with GenAI \cite{sarkar2023participatoryprompting,drosos2024rubberduck} and spreadsheets \cite{sarkar2020experience}. 22 participants reported a lot of experience with spreadsheets, and 2 participants reported they had some experience, but were still beginners. 15 participants reported they regularly used GenAI tools like Copilot or ChatGPT, with five having occasionally used, three casually tried, and one having not tried GenAI tools. Our relatively high proportion of GenAI users is due to the technologically-oriented skew of our participant pool, who are more likely to volunteer for studies of technology usage. Nonetheless, large scale surveys indicate that GenAI is rapidly increasing in adoption, with 39\% of the USA population aged 18-64 using generative AI in August 2024 \cite{bick2024rapid}. 11 participants were resident in North America, 6 in Europe, 4 in Africa, and 3 in Asia. Participants were compensated USD \$50 or local currency equivalent for their time.


\subsection{Procedure}
\label{sec:Methods:Procedure}
The study was approved by our institution's ethics and compliance board. Participants gave their informed consent during user study scheduling. 
The study was conducted remotely by researchers sharing control of the prototype through Microsoft Teams screen-sharing. Audio and video of the session were recorded and transcribed. 




Participants first completed a demographics form. Then, participants were given a tutorial using an example dataset showing how to upload a dataset, enter a prompt, interact with each element of the factor card, and view the shortlist being created.


When participants felt comfortable with the tool, the participant proceeded to the shortlisting task. The prototype was pre-loaded with a dataset, initial task objective, and initial factor cards. 

The participant was asked to inspect the cards and think aloud about their actions. This involved examining the factor title, selected column, factor criteria, level of importance, and, if applicable, the provocation. After inspecting a card, the participant decided whether to edit it, apply it to the shortlist, move to another card, or add a new card. 



Participants were asked to iterate until they were satisfied with shortlist, and to explain their final shortlist to the researcher. It was emphasised that participants should add, remove, modify, or apply factors based on their own personal rationale, and that there was no ``correct'' solution.


Half our participants saw provocations, and the other half did not (i.e., 12 in \ConditionA{} and 12 in \ConditionB{}). Due to the time intensive and deeply qualitative nature of our shortlisting task, each participant completed a single task during a 90 minute session (not including the tutorial task), to allow sufficient time for the briefing, tutorial, task, post-experiment interview and questionnaires. Task assignment was counterbalanced such that 6 participants in each condition performed Task 1 (Charity board) and 6 performed Task 2 (Movie night).

After the task, participants completed questionnaires and a semi-structured interview (detailed in Section~\ref{sec:Methods:DataAnalysis}).



\subsection{Data Collection and Analysis}
\label{sec:Methods:DataAnalysis}


We collected quantitative and qualitative data. Participants were asked to think aloud during their task. These utterances commonly focused on thoughts about each factor card and its elements (e.g., factor criteria, provocation, and importance), the temporary shortlist as factors were modified and applied, and goals for finalising the shortlist. 

Video recordings were analysed by two researchers who independently generated initial codes categorising participant events. They merged these into a single codebook of 12 specific events through negotiated agreement \cite{McDonald2019InterRater, saldana2021coding}. One researcher analysed all 24 recordings using this codebook to collect the frequency of each code occurrence. The results of this analysis are in Section~\ref{sec:Results:CODES}.



Participants in each condition completed at least three post-task questionnaires. The first two questionnaires assessed reflective thinking: one adapted from Kember et al.'s inventory \cite{kember2000development}, and a questionnaire on reflective thinking in AI workflows. Results of these are in Section~\ref{sec:Results:REFLECT}. The third questionnaire measured trust in automation \cite{jian2000foundations}, with results detailed in Section~\ref{sec:Results:TRUST}. Participants in \ConditionA{} completed a questionnaire on provocation effectiveness and user needs, with results in Section~\ref{sec:Results:ProvocationSurveys}. Questionnaires are reproduced in Appendix~\ref{Appendix:Instruments}.

Finally, we conducted semi-structured interviews about participants' experiences (questions in Appendix~\ref{Appendix:InterviewQuestionBank}). We performed a 3-phase thematic analysis \cite{Braun2006ThematicAnalysis} of participant interview answers and think-aloud utterances from the tasks:

\begin{enumerate}
\item Phase 1: One researcher first transcribed participant utterances, and applied iterative open coding \cite{Vaismoradi2013OpenCoding}, grouping quotes into an initial 4 theme groups containing 16 themes and 74 sub-themes.

\item Phase 2: A second researcher independently reviewed these coded utterances and evaluated each theme for the strength of its evidentiary support, discarding themes with weak support and recategorising utterances if deemed appropriate. This resulted in the refinement of the 4 theme groups into two main groups -- the effects of provocations on critical thinking (including 5 themes and 1 sub-theme), and the dimensions of the user experience of provocations (including 5 themes and 4 sub-themes).

\item Phase 3: Finally, through discussion, the two researchers aimed for negotiated agreement \cite{McDonald2019InterRater, saldana2021coding}. This resulted in refinement and agreement on 14 of 15 themes, with 1 theme discarded due to lack of agreement.
\end{enumerate}

The results of this analysis are in Sections~\ref{sec:Results:ProvocationEffects}~and~\ref{sec:Results:Dimensions}.

Our positionality as researchers invested in improving critical thinking creates a risk of bias towards favouring \ConditionA{}. To counteract this and improve the reflexivity of our thematic analysis \cite{braun2019reflecting}, we scrutinised our claims about \ConditionA{} to consider alternative and opposing interpretations. Thus, a large proportion of proto-themes from Phase 1 were discarded.

We achieved full (100\%) negotiated agreement and did not attempt to establish further inter-rater reliability with an independent third researcher, since our qualitative findings in Sections~\ref{sec:Results:ProvocationEffects}~and~\ref{sec:Results:Dimensions} do not hinge on frequency counts, and we are not intending to create a reusable codebook. The shared agreement established is sufficient to draw reliable conclusions about our specific dataset, aligning with the state-of-the-art guidelines for reliability analysis in CSCW and HCI research described by \citet{McDonald2019InterRater}. 

 

\section{Results}
\label{sec:Results}

Regarding \emph{RQ1: Do provocations affect users' critical thinking during shortlisting tasks?}, our qualitative analysis finds impacts on metacognition and critical thinking at multiple levels of Bloom's taxonomy (Section~\ref{sec:Results:ProvocationEffects}).

We did not find statistically significant effects of provocations on shortlist diversity (Section~\ref{sec:Results:RBO}), or the rate of think-aloud reflections and user interface actions (Section~\ref{sec:Results:CODES}). Though not statistically significant, \ConditionA{} had a higher median shortlist diversity and higher rate of think-aloud reflections.


We did not find significant effects of provocations on aggregate measures of self-reflection (Section~\ref{sec:Results:REFLECT}) or trust (Section~\ref{sec:Results:TRUST}).

Regarding \emph{RQ2: What important dimensions of the system design, task, or user interact with the effectiveness of provocations?}, we identify task urgency, task importance, user expertise, provocation actionability, and user responsibility as key elements contributing to critical thinking and the effectiveness of provocations (Section~\ref{sec:Results:Dimensions}).



\subsection{Shortlist Diversity With and Without Provocations}
\label{sec:Results:RBO}
\begin{figure}
    \centering
  \includegraphics[width=\textwidth]{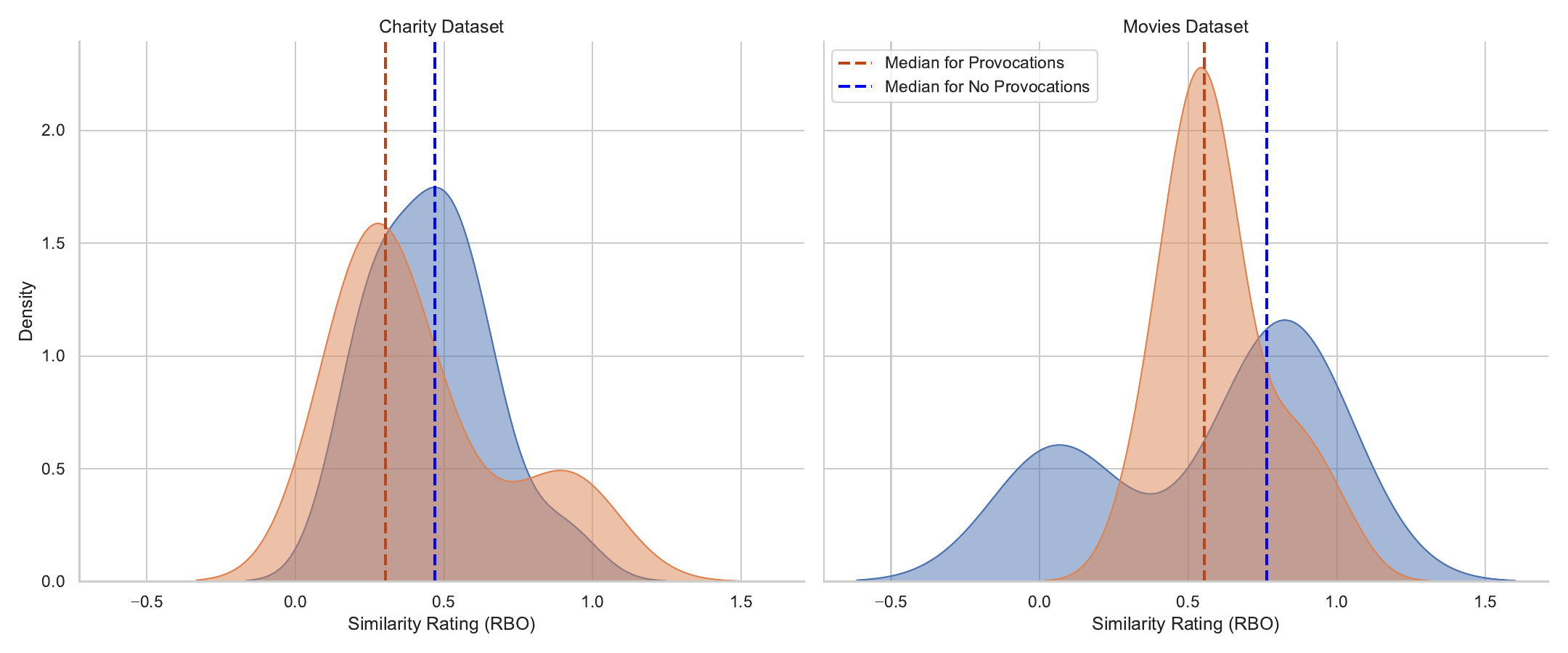}
  \caption{Distribution of final shortlist similarity for the four groups of conditions and tasks within the study (Charity task on the left, Movies task on the right). This kernel density estimate (KDE) plot~\cite{seabornKDE} visualises the Rank-Biased Overlap (RBO) similarity metric, where a higher RBO means more similar lists. Median RBO values for each dataset in \ConditionA{} (in orange) are lower than the respective values for \ConditionB{} (in blue).}
  \Description{}
  \label{fig:RBO}
\end{figure}



We compare the similarity of participants' final shortlists using Rank-Biased Overlap (RBO)~\cite{Webber2010RBO,rboPypi}, a list similarity metric where scores closer to 1 indicate higher similarity (1 = identical lists, 0 = completely different). Participants fall into 4 groups by task and condition: \ConditionA{} Charity, \ConditionB{} Charity, \ConditionA{} Movies, and \ConditionB{} Movies. We computed RBO for each pair of participants within each group, resulting in 15 scores per group. Figure~\ref{fig:RBO} shows the distribution and median of these scores.


In the Charity dataset, \ConditionA{} had a lower median RBO (0.30) than \ConditionB{} (0.47), and similarly for Movies (\ConditionA{}: 0.55 vs. \ConditionB{}: 0.77). Thus, \ConditionA{} had higher solution divergence in both datasets, although the difference in distributions is not statistically significant (Mann–Whitney U tests).

\subsection{Frequency of Think-Aloud and User Interface Actions With and Without Provocations}
\label{sec:Results:CODES}




\input{Tables/tbl_actioncodes}

We coded participants' think-aloud utterances and interface actions into the categories shown in Table~\ref{tbl:actioncodes}. Think-aloud codes capture participant utterances about the interface. Action codes represent participant interactions with the system during the task.





\begin{figure*}
    \centering
  \includegraphics[width=0.45\textwidth]{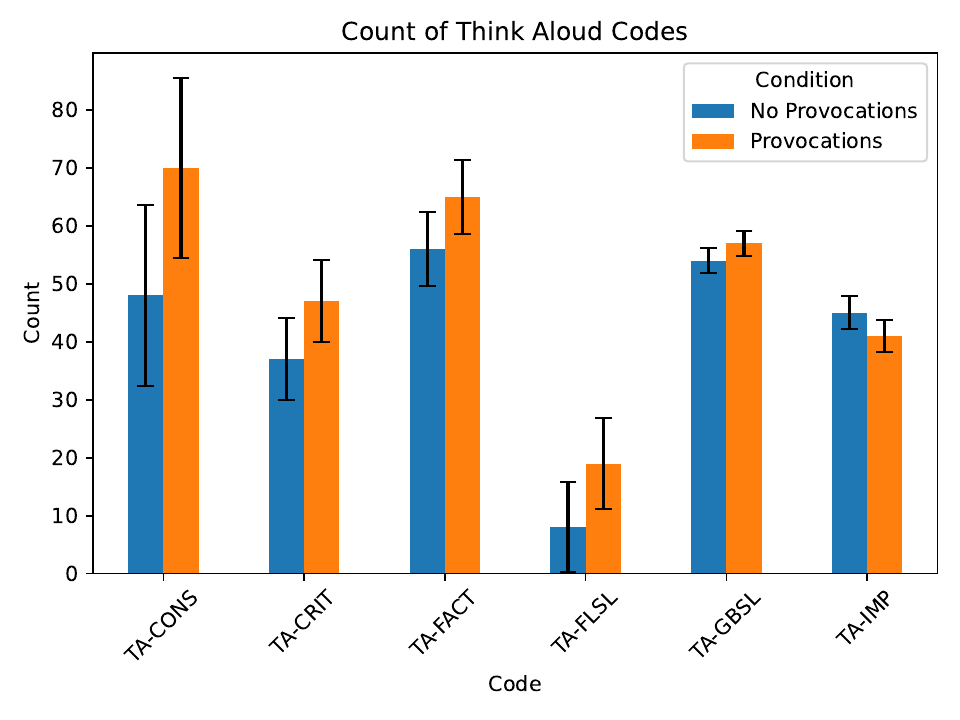}
  \includegraphics[width=0.45\textwidth]{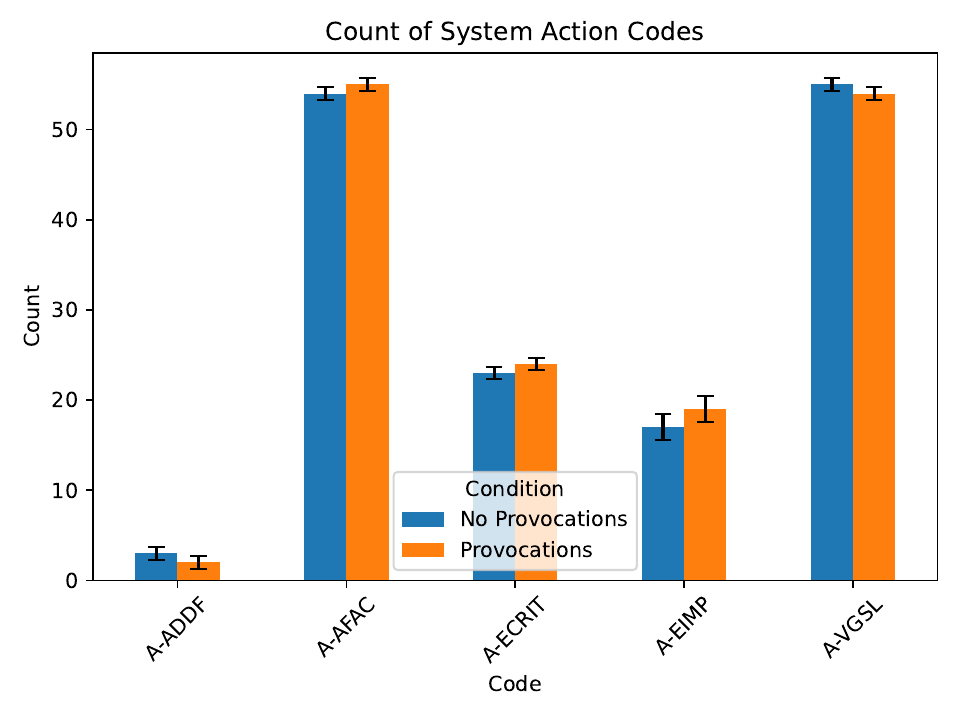}
  \caption{Occurrence counts of think-aloud codes (left) and action codes (right) per condition. On average, there were more occurrences of think-aloud codes in \ConditionA{}, but a similar number of actions for both conditions. Code definitions in Table \ref{tbl:actioncodes}.}
  \Description{}
  \label{fig:thinkalouds_action_codes}
\end{figure*}


\ConditionA{} had a higher occurrence of all think-aloud categories except for TA-IMP (reads text from or mentions factor importance), but these differences are not statistically significant (Chi-Square tests)(Figure~\ref{fig:thinkalouds_action_codes}). Participants in \ConditionA{} performed 155 actions, versus 152 for \ConditionB{}, which is not statistically significant (Chi-Square test).

\subsection{Self-Reported Reflection With and Without Provocations}
\label{sec:Results:REFLECT}
\subsubsection{Kember Reflective Thinking Inventory}
\label{sec:Results:KEMBERREFLECT}


We compute a ``Reflective Thinking Score'' by averaging responses to the 16-question Kember et al. \cite{kember2000development} Reflective Thinking inventory (with minor adaptations, reproduced in Appendix~\ref{Appendix:ReflectiveThinkingSurvey}). The median score for \ConditionA{} was 3.31 versus 3.59 for \ConditionB{}; the difference is not statistically significant (Mann-Whitney U test).



Within the inventory, two questions had significant differences (Mann-Whitney U test). For the question: ``With this tool, the shortlist task requires me to understand the domain knowledge of the dataset used for shortlisting'',  \ConditionA{} has a significantly lower median of 4.0 (``Agree with reservation'') versus 5.0 (``Definitely agree'') for \ConditionB{} ($U=36.0, p=0.023$). For the question: ``With this tool, I need to understand the dataset and the shortlist criteria in order to perform practical tasks'', \ConditionA{} has a significantly lower median of 4.0 (``Agree with reservation'') versus 5.0 (``Definitely agree'') for \ConditionB{} ($U=40.0, p=0.045$). This indicates that provocations can scaffold shortlisting tasks in unfamiliar domains (supported by think-aloud remarks), while introducing the potential for overreliance.





\subsubsection{AI Workflow Reflective Thinking}
\label{sec:Results:WORKFLOWREFLECT}





Pilot studies revealed that participants struggled to relate the questions in the Kember et al. Reflective Thinking inventory to their concrete experiences with the prototype. Therefore, we developed a \emph{de novo} questionnaire to capture reflective thinking behaviours with AI-assisted workflows, modelled after Kember et al. (detailed in Appendix~\ref{Appendix:SystemImpactSurvey}).


We derive a participant ``AI Workflow Reflective Thinking'' score by averaging responses to this 14-question questionnaire. The median score for Condition A (Provocation) was 3.68 (mean=3.70) versus 3.79 (mean=3.83) for Condition B (No Provocation), and this difference is not statistically significant (Mann-Whitney U test). Within the inventory, there is a significant difference for the question: ``I considered the possibility that the AI suggestion could be wrong''. The median for \ConditionA{} was 3.5 (between ``Neutral'' and ``Agree'', mean=3.08) versus 4.0 (``Agree'', mean=4.08) for \ConditionB{} ($U=39.5, p=0.039$). This result is surprising, and indicates potential overreliance on provocations.




\subsection{System Trust With and Without Provocations}
\label{sec:Results:TRUST}
System Trust was measured using the Checklist for Trust between People and Automation inventory \cite{jian2000foundations}. The median score for \ConditionA{} was 5.29 (mean=5.30) versus 5.04 (mean=4.88) for \ConditionB{}, which is not statistically significant (Mann-Whitney U test). Within the inventory, there is a significant difference between conditions for the question ``The system provides security'' ($U=122.5, p=0.003$). The median value for \ConditionA{} was 5.0 versus 3.0 for \ConditionB{}, suggesting that provocations helped participants feel more secure and comfortable about AI suggestions.

\subsection{Provocation Evaluation Questionnaire}
\label{sec:Results:ProvocationSurveys}

\begin{figure}
    \centering
  \includegraphics[width=0.80\textwidth]{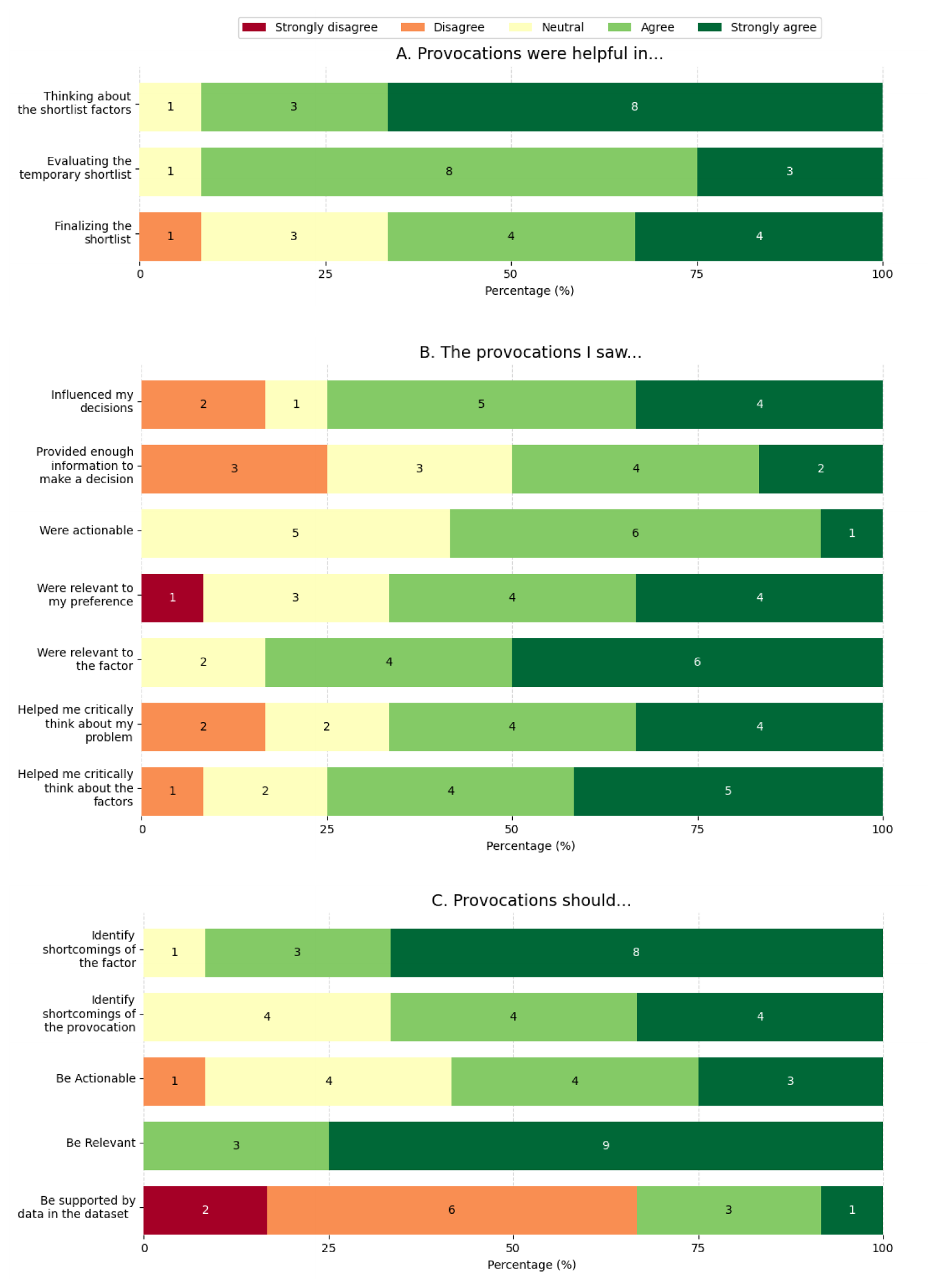}
  \caption{The results of a Provocation Evaluation Questionnaire measuring \ConditionA{} participant agreement rating from Strongly disagree to Strongly agree. Numbers within the bars of each section are the count of participants.}
  \Description{}
  \label{fig:COMBINED}
\end{figure}

Participants in \ConditionA{} also completed a 15-statement questionnaire consisting of 5-point Likert items from ``Strongly disagree'' to ``Strongly agree'' about their perceived needs for and impacts of provocations. Their responses are given in Figure~\ref{fig:COMBINED}.


Overall, participants agreed that provocations were helpful for thinking about the shortlist factors (i.e., the factor cards), evaluating the temporary shortlist and finalising the shortlist (Figure~\ref{fig:COMBINED}, A). 

Participants agreed that provocations were relevant to the factors and their preference (i.e., the initial prompt and goal). They agreed that provocations influenced their decisions, that provocations were actionable (e.g., they knew the next step required to apply the provocation), and that provocations helped them critically think about the factors and their overall problem. They were divided about whether provocations provided enough information to make a decision (Figure~\ref{fig:COMBINED}, B). 

Participants strongly agreed that provocations should identify the shortcomings of the factor (e.g., explain why a factor might not be suitable), and agreed that provocations should identify shortcomings of the provocation itself. Participants also strongly agreed that provocations should be relevant to the dataset or user goal, but were divided on whether provocations should be actionable. Participants tended to disagree that provocations should be only supported by data in their dataset instead of leveraging external information (e.g., supplementary datasets, relevant contextual information from the Web or other user files) (Figure~\ref{fig:COMBINED}, C).

\subsection{Qualitative Effects of Provocations on Critical Thinking}
\label{sec:Results:ProvocationEffects}

With some exceptions, our quantitative comparisons are not statistically significant. This is likely due to a combination of the inadequacy of our quantitative measures, insufficient statistical power, and the relative infrequency of conscious critical thinking ``events'' (discussed in Section~\ref{sec:discussion}). However, this obscures the qualitative effects on critical thinking observable in the think aloud data.

In this section, we describe how provocations resulted in consciously articulated critical thinking events. We coded participant utterances according to the ``level'' of critical thinking they corresponded to in Bloom's taxonomy \cite{bloom1956taxonomy,huitt2011bloom}: analysis, synthesis, evaluation, comprehension, and application. Additionally, we observed a number of utterances corresponding to increased metacognition. Research has uncovered a range of metacognitive demands associated with GenAI \cite{tankelevitch2024metacognitive}. Metacognition and critical thinking are related; some critical thinking processes are primarily metacognitive in nature (e.g., reflecting on the limits of one's own knowledge) and we find it more useful to characterise them as such.


We note that in participant utterances, there is not always a neat separation between the levels of Bloom's taxonomy, or between Bloom's taxonomy and metacognition. Many participant utterances reflected multiple levels simultaneously, or reflected the fluid and instantaneous transition from one type of thinking to another. For simplicity, we focus on one level at a time, but recognise that this entails some degree of artificial separation.

The results in this section are drawn only from participants in the with-provocations condition. For brevity, the comparison to participants in the without-provocations condition is implicit; we present only those consciously articulated examples of effects on critical thinking which occurred exclusively in the with-provocations condition.




\subsubsection{Metacognition}

Exposure to provocations triggered a range of metacognitive processes. Provocations that highlighted some aspect of a factor or nuance of the problem that the participant had overlooked caused participants to consider what else they might have missed. For example, provocations assisted in P1's current strategy of \emph{``thinking about an alternative, [...] produce the data, then look at it and think about what did I miss instead?} Provocations caused participants to be \emph{``primed to think of other things to consider''} when considering factors (P1).

\paragraph{Mental Model of AI}

Provocations triggered metacognitive reflections about the user's mental model of the AI, their assumptions about the system's operation, and created opportunities to update them. Provocations gave participants ideas for \emph{``changing the prompts''} (P13) and \emph{``helps you with that prompting''} (P1), and P1 thought GenAI was \emph{``only as good as your ability to prompt it.''} Participants also reflected evaluatively about the capabilities of the model, which influenced their problem solving. P21 \emph{``needed to pay attention to''} provocations to understand the \emph{``possibility that a response could include or wrongly generate new information''}.



Participants' mental model of the system interplayed with their perception of task urgency (detailed in Section~\ref{sec:Results:Dimensions}). 
If a user felt \emph{``under pressure''} to complete a task quickly, \emph{``you're not going to spend time messing around with something you're not 100\% familiar with''} and instead would \emph{``default back to your own skill set''} (P1).
Participants stated that further experience with provocations would impact their use of the system, \emph{``because you would get more familiar''} (P7) with the system over time.


%


\subsubsection{Bloom Level: Analysis}
Analytic thinking is concerned with defining a problem, such as by decomposing it into its constituent parts, considering how those parts interrelate, and making logical inferences. Provocations triggered analytic thinking in multiple ways.

Provocations caused participants to consider their problem in greater detail, reflect on and refine their objectives, break down the problem into sub-problems to gain clarity, and consider the implications of their varying approaches.
P9's provocations led them to consider \emph{``recommendations for additional data columns''} not present in the dataset. P1's provocations helped them understand the need to \emph{``cast the net a bit wider''} and add more keywords to a factor.

Provocations also triggered analytic thinking about the dataset, helping the user understand the limitations of their dataset, and think about approaches to solving their problem more broadly than the data in front of them. Even when participants judged a provocation as irrelevant, it still served to induce analytic thinking about their problem because it forced them to articulate more clearly what was, in fact, relevant. P11 thought the provocation critiquing box office performance was \emph{``not useful''}, but only after reflecting on \emph{``what actions [... would] go deeper into''} that issue. P5 thought such provocations were \emph{``interesting [... for] scoping out the amount of research that would be required to assess''} the problem further.


\subsubsection{Bloom Level: Synthesis}


Synthetic thinking involves ideation and the combination of existing ideas to create new ones. Synthetic thinking triggered by provocations typically took the form of developing strategies for approaching the problem in response.

Analytic and synthetic thinking were often induced in succession by provocations. Provocations caused participants to analytically refine or decompose their problem, before synthesising solutions.
A provocation prompted P21 to consider missing factors they had \emph{``not previously thought about''}, and it \emph{``certainly gave insights for shortlisting the dataset''} resulting in the decision to add \emph{``another factor to narrow down that data set more.''} 
P1 viewed provocations as a problem-solving resource that would \emph{``bring far more intelligence''} beyond the current dataset.

\subsubsection{Bloom Level: Evaluation}
Evaluative thinking is the exercise of critical judgement, using and applying criteria to judge statements, artefacts, and ideas.



Participants engaged in evaluative thinking by assessing their problem as well as the provocations, and developing and applying internal criteria to judge the relevance of provocations to their context. For example, P17 compared a provocation that was \emph{``warning that (the factor) may exclude grants that benefit children indirectly''} to a card game where playing certain cards is technically permissible but may not have any effect, suggesting that P17 was evaluating provocations along the dimensions of relevance, correctness, and actionability.

Evaluative thinking often co-occurred with analytic thinking and metacognition. A provocation might first cause participants to deconstruct their problem (analytic), reflect on the limitations of their own knowledge (metacognitive), and then judge the quality or importance rating of a particular factor (evaluative). For example, P7 thought the provocation around Rotten Tomatoes scores was useful \emph{``because it explains the basis for doing those ratings''}, causing them to update the importance of the factor since it came from critics who \emph{``would know rather more about the film business''} than they did. Similarly, P5 considered which factors were an actual \emph{``assessment of quality''} and thought \emph{``some of the other factors were a better signal for quality''} than box office performance, since the provocation highlighted that box office performance reflected \emph{``the popularity of the movie, and less likely [its] actual quality.''}

\subsubsection{Bloom Level: Comprehension and Application}
Provocations can cause thinking also at the ``lower'' levels of comprehension (developing an understanding of facts and concepts) and application (applying knowledge learned in one context to another). Provocations helped P23 \emph{``understand what box office performance actually measures''}, and reflect on \emph{``how many people actually gave the movies Rotten Tomatoes scores''}, consequently decreasing its factor importance.

Provocations operationalise statements about the data in ways that expose the user's limited understanding of it. If a participant fails to understand a provocation, it can trigger the realisation that they do not fully understand the data. Provocations take otherwise ``inert'' data about which participants may have made assumptions, and couch them in statements that expose these assumptions. For example, when P19 did not fully comprehend what ``distribution'' meant within the context of box office performance, and the provocation was \emph{``a really helpful start''} for exploring what the term \emph{``might mean''} by \emph{``grounding the terms or connections that (the provocation) is trying to make with more concrete ways that we might think about these terms''}.


\subsection{Dimensions of the User Experience of Provocations}
\label{sec:Results:Dimensions}


Apart from direct effects on critical thinking, we also observed certain dimensions along which the user experience of provocations varied. These include: task urgency, task importance, user expertise, provocation actionability and provisionality, and user responsibility. These are particularly salient aspects of the user or task that appear to mediate participants' relationship with and response to provocations. They are not design dimensions, i.e., neat continua that can be manipulated through design choices. However, we can draw some design implications (Section~\ref{sec:discussion}).

\subsubsection{Task Urgency}
\label{sec:Results:Dimensions:TaskUrgency}

Participants reflected that their attitudes to provocations would vary depending on whether the participant was under time pressure. Provocations could \emph{``put someone off if they are doing a quick and dirty job''}, but were \emph{``very useful for a large project or exploratory search of data''} (P13).


Participants commented on the iterative and continuous nature of provocations. There is no objective guide for when ``enough'' critical thinking has been done, and for each new idea developed by the participant, there always remains scope for further provocation. 
P1 thought users \emph{``should be massively encouraged to run (the system) a few times and put in different criteria to see what you come up with, and refine the data set and the ranking.''}
The subjective judgement calls made by participants to declare that ``enough'' attention had been paid were influenced by many of the dimensions we discuss here, but task urgency was most frequently cited.



\paragraph{Warning Fatigue}
We observed the potential for warning fatigue (similar to alert fatigue \cite{cash2009alert}). Participants compared provocations with a class of warnings, such as end-user licence agreements, cookie consent notifications, etc. which are routinely ignored. If provocations are viewed to fall in this category, participants may ignore them, especially when the task is urgent. P15 said they were \emph{``more likely to just ignore''} provocations that were not actionable, as they were \emph{``used to seeing fluff (warning) text and ignoring it.''} Provocations that were \emph{``not ambitious enough''} (P13) or \emph{``stating the obvious''} (P23) added to this perception. Provocations needed to \emph{``challenge or widen how someone might perceived the data''} to be valuable (P9).

Besides making provocations less generic, a clear design implication is to not present provocations using the conventional visual language of alerts, warnings, and notifications. Rather, provocations should be presented in a manner that suggests they are part of the user's tool or workflow. This is an opportunity for future work.



\subsubsection{Task Importance}

Participants stated that their propensity for attending to provocations would depend on task importance, expressed in terms of its external consequences (i.e., whether it was ``high stakes'' or ``low stakes''), and the participants' internal attitude of seriousness or commitment to the task.
For example, for \emph{``very low stakes task, like shortlisting movies''} they \emph{``might do (the task) without thinking too much''}, but for tasks like \emph{``giving somebody a grant of thousands of pounds''}, they \emph{``would like to think more throughout the process, ideally''} (P17).

The movie night task was seen as relatively unimportant, as the consequences of this shortlist were minimal, and participants adopted a casual attitude to it. 
P5 stated the task had \emph{``the lowest stakes possible as there's really no risk associated with it.''} 
In contrast, the charity task admitted a much wider variety of interpretations and consequent attitudinal dispositions. It was seen as a \emph{``higher stakes domain''} where decisions \emph{``could potentially be harmful''} (P5).
Depending on the extent to which participants chose to engage realistically in the task, they reflected on the consequences of allocating grant funding and cited the seriousness of the task as being a contributing factor in their attendance to provocations.
Similarly, P15's trust in the system depended on task importance, and for tasks that their \emph{``boss might see''} it was important that they \emph{``double check''} AI suggestions.



\subsubsection{User Expertise}

The user's domain expertise also mediated their experience of provocations. 
For example, P5 \emph{``appreciated that the model is bringing forward some of these things that, as a non domain expert, I wasn't really considering''} when offered a provocation to the \emph{Past Performance} factor, as they had not considered that newer charities may not have had the opportunity to build a track record of performance.
Participants with domain expertise were accustomed to thinking critically about similar problems, so provocations could be \emph{``just a reflection''} of participants' prior knowledge, and \emph{``did not change or improve''} their thinking (P9). Thus, to be useful to domain experts, there is a need for provocations to provide more detailed, counterintuitive challenges, and avoid telling the user what they likely already know.

User expertise in generative AI may impact user needs around provocations, but we did not observe any strong indications for this. 
Participants P15 and P19 were the only two in \ConditionA{} that replied to the demographics questionnaire that they only \emph{``casually tried''} GenAI tools. P19 prioritised autonomy and desired not being told exactly what to do by provocations (Section~\ref{sec:Results:Dimensions:ActionabilityProvisionality}), and P15 would ignore non-actionable provocations and preferred to be able to quickly apply provocations they deemed useful (Section~\ref{sec:Results:Dimensions:TaskUrgency}). 
These participants were also the only ``Disagree'' responses for the question ``The provocations I saw influenced my decisions'' (Figure~\ref{fig:COMBINED}). Our sample of casual users is too small to allow us to generalise beyond these specific participants. 
Beyond these two observations, we did not see differences between this group and participants with greater GenAI use.



\paragraph{Comprehension for Provocations}

We observed cases where explanations, examples, inline definitions, and justifications would help users understand provocations.

Participants suggested displaying example rows where two factors might have mutually exclusive criteria (P17) and show \emph{``visualisations on how different fields in the dataset correlate''} (P13), which might include \emph{``word clouds''} with \emph{``commonly associative words''} to assist in keyword selection (P7).
They also sought greater justification and explanation for provocations. For example, P15 wanted the provocation to \emph{``just show what's wrong''} using example rows where a factor might not actually match the user's goal, because \emph{``otherwise it's just noise.''} 
P13 wanted a list of \emph{``fields that weren't used''} and help to understand why they were not used as potential factors \emph{``to explain to an analyst why they got ignored and as a sanity check that they have not left something out, because if I'm talking to a client and I run their data set through this, and it doesn't use their favorite field, they'll want to know why.''}

\subsubsection{Provocation Actionability and Provisionality}
\label{sec:Results:Dimensions:ActionabilityProvisionality}
As seen in their questionnaire responses (Section~\ref{sec:Results:ProvocationSurveys}) participants were divided on whether provocations should be ``actionable''. Some provocations offered a direct suggestion that could be actioned within the tool, e.g., suggesting adding certain keywords to the description of the factor, and there was a clear design opportunity to help the user incorporate this suggestion, e.g., by offering a button to add the keywords, rather than expect the user to manually edit the description.



However, many provocations were not so direct. In some cases, the provocations suggested actions to take outside the tool or the users' present workflow. These were still in some sense ``actionable''. For example, P13 thought that despite a provocation not being clearly actionable, it was still \emph{``helping out with what to say and what to ask (the AI) to do next''}.




Our analysis suggests a hierarchy of levels of actionability, organised by increasing cognitive and interactional effort: 
\begin{enumerate}
\item Apply without asking: the least effort, the participant does not have to take any action to apply the provocation's suggestions, it is done automatically. 
\item Click to apply: the suggestion can be actioned within the tool, and the tool automates this action upon the user's request, e.g., \emph{``a button that says do this action''} (P15). 
\item Direct in-tool user action: the suggestion can be actioned within the tool, but the user carries out the interaction steps necessary. 
\item Direct out-of-tool user action: the suggestion can be actioned, but it requires the user to leave the tool and immediate workflow, e.g., retrieve another column of data that is not present in the dataset. 
\item Indirect user action, whether in-tool or out-of-tool: the suggestion does not directly suggest a follow-up action, but it prompts reflection that in turn suggests an action, e.g., the user reads a provocation and realises they do not understand the meaning of a particular term, thus decides to take action by revisiting some documentation.
\item Reflection: the suggestion offers no follow-up action, directly or indirectly, but causes the user to reflect on their own understanding and makes them consciously aware that they don't know how to proceed.
\end{enumerate}



To address warning fatigue, provocations at different levels of actionability could be presented differently. For instance, while it is difficult to separate direct from indirect actionability (the indirect nature relies on user reflection, and thus cannot be detected by the tool itself), provocations that contain direct, in-tool actionable suggestions could be presented differently from those that don't.

\paragraph{Provisionality}

As a counterpart to actionability, participants also experienced provocations as having varying degrees of \emph{provisionality}, which can be expressed as follows: provocations should offer a \emph{potential benefit}, but not \emph{compel} the user to respond to it. P19 enjoyed provocations suggesting things to consider \emph{``without saying `you must do this'''}.
Provisionality is analogous to receiving feedback from colleagues. When one solicits feedback from a colleague on some work, one is expecting remarks and critiques drawing on the colleague's perspective and expertise that could potentially benefit the work, yet one is not compelled to agree with, respond to, or implement every aspect of this feedback. 

Participants experienced what can be described as \emph{productive confusion}, leaving them \emph{``with more questions''} than they had before, which was seen as \emph{``not necessarily a bad thing''} for P5 as it led them to having to take a step back from interacting with the UI to consider how (or if) a factor should actually be applied. 
Provocations induced confusion in participants -- sustained episodes of high cognitive effort that do not culminate in a clear, outwardly visible decision. 
Yet this confusion can still be productive and desirable.
Participants remarked that the confusion induced by provocations helped to push them out of their comfort zones.
For example, P11 initially considered box office performance as \emph{``an important factor''}, but being confused by the provocation lessened the perceived importance of the factor for them and instead considered other factors more heavily. While they initially viewed the provocation as not actionable, it nonetheless spurred them to think analytically about the problem and resulted in a productive change in behaviour.

Productive confusions can be productive by inducing metacognitive thinking. For example, P17 initially saw one provocation as \emph{``just a warning and not necessarily a tip for how to change the criteria.''} Later, P17 was \emph{``thinking about the criteria,''} and how it could be applied to specific rows. Subsequently, the provocation \emph{``even help[ed] counter biases''} they had, namely, assuming a linear relationship between grant size and impact.


Rather than respond directly to provocations, participants often responded to provocations by shifting their general reflective practices; in such cases provocations functioned as ambient, peripheral, almost subconscious aids to reflection.


\subsubsection{User Responsibility}
Participant responses to provocations were affected by how responsible or accountable they felt in the process of making decisions with AI assistance.

AI assistance complicates participants' understanding of their responsibility. The story of P9 is illustrative. P9 claimed that provocations allowed them to think less about structuring their data -- a type of analytic thinking, since data structuring involves decomposing a problem into parts that can be mapped onto affordances such as spreadsheet columns \cite{chalhoub2022freedom}. However, the provocations in fact caused them to think more about structuring their data; a provocation that suggested adding a column triggered analytical reflection about the suitability of the column. We hypothesise that having to ``think less'' in this instance indicates how provocations allowed the user to feel less ``responsible'' for considering edge cases and possible additions.

Different participants felt responsible and accountable for different stages of the process. Some thought it was best to intervene at the final shortlist stage, and they \emph{``would find it hard to resist going back over the entire list to see nothing has been overlooked, because of something the AI didn't ask me''} (P7). P5 desired a \emph{``summarization''} provocation for the final shortlist, based on the considered factors, to help evaluate if a user met their goal. Others speculated that other stages were most appropriate for human intervention.
Some thought oversight over each factor and provocation \emph{``depended on the user interacting with the system''} (P9) and on the complexity of the dataset (P19).
P7 considered potential negative consequences if certain charities were not shortlisted because \emph{``they did not use certain keywords in their description''}, and wanted to consult human collaborators about factors and criteria they were considering.



Participants revealed a tension between wanting to think, and wanting not to think. 
Or between wanting to think, and not wanting to think too much, since users \emph{``would be disheartened to use this tool''} if there was an overwhelming amount of factors to consider (P23). P23 described losing track of each decision they made as they considered each factor and provocation, worried that they \emph{``did not weigh the factors against each other well''} due to information overload.
Indeed, one of the value propositions of GenAI tools is that they can perform the difficult, burdensome, or tedious activities of cognition -- to spare the user from having to think, in other words. 
On the other hand, participants, through their concern for responsibility, also wished to do some thinking for themselves.
P23 thought one solution might be to only show factors (and provocations) that are considered high or critical importance, which may be where users experience the greatest need to express their autonomy and responsibility in the decision-making process. 
The tension between wanting to think and not to think is affected by the dimensions of task urgency and importance: urgent tasks afford less time for thinking, and important tasks demand more time for thinking.

Responsibility relates to agency (loosely defined, the perception of one's ability to effect change in the world \cite{coyle2012did}) and autonomy (loosely defined, the perception of one's ability to act without support or influence from other agencies). 
For P19, provocations using the word \emph{``may''} was \emph{``exactly what [provocations] should be''} and not \emph{``doing the decision making for you''}. 
P9 saw provocations as \emph{``recommendations''} that could be overridden by the user's \emph{``personal preferences''} since they still \emph{``had the autonomy''} to make a final decision.
P17 thought this autonomy valuable, as even if AI-generated solutions \emph{``might not be the best, for shortlisting it might still be fine if the final decision is being made by a group of people.''}


\section{Discussion}
\label{sec:discussion}

\subsection{Connections to Related Work}

The way in which provocations can introduce ``productive confusions'' (Section~\ref{sec:Results:Dimensions}) is related to a family of design research that subverts seamlessness and direct utilitarianism (e.g., \cite{chalmers2003seamful,gaver2003ambiguity,rossmy2023point,sarkar2023simplicity}). It has been proposed that AI should act as a provocateur \cite{sarkar2024challenge,sarkar2024copilot}, act as an antagonist \cite{cai2024antagonistic}, act as a cognitive forcing function \cite{bucinca2021trust}, cause cognitive glitches \cite{hollanek2019non}, or involve attention checks \cite{gould2024chattldr}. More generally, provocations can be seen as an instance of a microboundary, or design friction \cite{cox2016design}. Provocations lend a new dimension to this area: namely, that such frictions can be dynamically and contextually generated with the specific aim of improving human critical thinking. Productive confusions appear to be related to the pedagogical concept of ``productive failure'' \cite{kapur2008productive} and future work may investigate this connection.

Provocations are related to ``nudges'' \cite{thaler2008nudge}, a term from behavioural economics meaning \emph{``any attempt at influencing people’s judgment, choice or behaviour in a predictable way which works by making use of [people’s] boundaries, biases, routines and habits as integral parts of such attempts''} \cite{congiu2022review}. Provocations too aim to influence people's behaviour, but in a more abstract manner than nudges, in that they aim to induce deliberative decision-making and thereby expand the user's freedom of choice, countering the tendency for mechanised convergence. In contrast, nudges are deployed to induce specific behaviours that the so-called ``choice architect'' \cite{sunstein2017nudges} deems to be desirable, such as smoking less, exercising more, or eating healthier foods. In a traditional nudge, confusion is undesirable and may result in the nudge being ineffective \cite{sunstein2017nudges}, whereas for provocations, confusion can be productive (Section~\ref{sec:Results:Dimensions}).

Provocations relate to ``reflection prompts'' or ``metacognitive prompting'' in education (e.g., \cite{salomon1988ai, bannert2006effects}), and HCI research (reviewed in Section~\ref{sec:related-work-design}, for an arbitrary example of the term in use see \citet{roldan2021pedagogical}). In HCI, the term ``reflection/reflective prompt'' is used to mean any stimulus that aims to promote reflection on any topic, from mental health to misinformation. In education, it specifically denotes stimuli that promote metacognition about one's learning aims and progress. Provocations can be considered a type of reflective prompt, with the following specific characteristics: (1) provocations are stimuli that aim to induce critical thinking as defined by a framework such as Bloom's taxonomy, (2) provocations are applied in the context of AI-assisted knowledge workflows where AI assistance poses a risk to critical thinking.


Provocations also relate to concepts such as extended mind, extended cognition, external cognition, and distributed cognition (see, e.g., \cite{malafouris2019mind,turner2016distributed,rupert2004challenges}). These posit that thinking does not happen only in the user's mind, but rather is distributed across multiple people, tools, and artefacts. Even in the single user setting of our study, thought is enacted across the user, the shortlisting tool, the AI assistance elements of the shortlisting tool, and the AI provocation elements of the shortlisting tool. As observed in Section~\ref{sec:Results:Dimensions}, users varied in their conceptualisations of the loci of responsibility across the workflow. Similarly, previous work finds that the ambivalence of users towards their accountability is a key aspect of the social construction of datasets \cite{orr_crawford_2023}. If critical thinking is viewed as a property not just of the user but rather of the system comprised of the user, their tools, their artefacts, and their AI agents, it may be possible to design for \emph{distributed critical thinking}. This raises the question whether, in terms of a design goal, it is sufficient to consider only the quality of the final artefact produced (such as a shortlist), as opposed to outcomes in the user's mind. That is, if the final artefact is high quality, and critical thinking occurred \emph{somewhere} in the system, does it matter whether it improved critical thinking in the user's mind? We propose that it does matter. As Kidd observed, the most important aspect of knowledge work is the private transformation of the knowledge workers' thought; the artefact is incidental \cite{kidd1994marks}. As such, even in a distributed cognition regime, we believe that critical thinking design interventions should aim to improve the critical thinking performed by the user, and develop their enduring skill in doing so.

\subsection{Difficulties in Quantifying the Impact of Provocations on Critical Thinking}
We observed deep qualitative differences in participants' critical thinking behaviours with provocations (Section~\ref{sec:Results}). However, few of our quantitative comparisons were statistically significant. There are multiple possible reasons for this. First, as described later in Section~\ref{sec:limitations}, our study is underpowered to detect effects of the size we observed. In future work, a larger sample will be necessary to detect quantitative effects (yet it may not be sufficient, because of the following two reasons, which will also need to be addressed).

Second, we found that despite using validated, state-of-the-art questionnaires for reflective thinking and trust, these questionnaires are still difficult for participants to interpret. Future work should develop new self-reported measures of critical thinking, and explore alternative proxies for critical thinking (e.g., EEG to measure cognitive load \cite{antonenko2010using}).

Third, any effects of our critical thinking intervention occur against the backdrop of an individual's pre-existing critical thinking skill and disposition, and general cognitive demands of the shortlisting task that are not specific to provocations. Thus the magnitude of any shifts in critical thinking due to provocations must be so substantial that they would ``show through'' these other influences in a between-subjects design. Finally, many effects on critical thinking were counterintuitive (e.g., ``productive confusions'', Section~\ref{sec:Results:Dimensions}) and unlikely to be consciously recalled in a self-report questionnaire.


\subsection{Design Implications for Provocations}

Recall that our primary aim is to understand the human experience of critical thinking aids in the context of AI-assisted workflows. Nonetheless, our findings suggest some strategies for improving the effectiveness of provocations:

\emph{Provocations should aid comprehension}. 
Some participants thought a provocation was not useful because they could not interpret it, or did not know how to apply its suggestion.
Through explanations, provocations could assist users in understanding \emph{why} they should apply it, or allow users to interrogate a provocation to learn about what it addresses, and \emph{how} to take action in response.

\emph{Provocations should be relevant, but not redundant}. 
Several participants noted that some provocations were redundant. For example, the provocation to the Family Genre factor told users to also consider the View Rating, which was already being considered. 
While this could still be useful for helping a user understand the importance of a factor, provocations should be aware of other provocations and data in the workspace.


\emph{Provocations could go beyond text}. Several participants (P7, 13, 15, 19) noted that visualizations might improve provocations. Text may be seen as ``legal jargon fluff'' (P15) similar to other warnings shown with AI-generated content. For our shortlisting tasks with spreadsheet data, participants suggested visualizing how data correlated with each other.



\emph{Provocations could be interactive}. Our static text provocations rely on users to read, consider, and apply them. Interaction, even a simple accept or deny, or presenting provocations as an interactive chat dialogue, could improve engagement and discourage ignoring them \cite{chi2014icap}.

\section{Limitations}
\label{sec:limitations}
We conducted a power analysis across small, medium, and large effect sizes (Cohen's $d$ = 0.2, 0.5, and 0.8, by convention). With 12 participants per condition, our power for large effects is 0.47. Thus, our study was underpowered to consistently detect quantitative effects in the questionnaires (Section~\ref{sec:Results:REFLECT} and Section~\ref{sec:Results:TRUST}). 
For our questionnaires we calculated observed effect sizes of d=0.45 (for the analysis in Section~\ref{sec:Results:KEMBERREFLECT}), d=0.49 (for Section~\ref{sec:Results:WORKFLOWREFLECT}), and d=0.49 (for Section \ref{sec:Results:TRUST}). These are moderate effects in practice, but our study is underpowered to detect them. The minimum effect size for which our sample reaches the conventional target power of 80\% is 1.2, larger than the effects we observed. For 80\% power in detecting large effects, we require approximately 26 participants in each condition (52 overall). This would have resulted in a prohibitive  increase in the cost and complexity of recruitment, data collection and analysis.

With the Charity dataset task, two of the factors generated were not attached to any columns in the dataset. While not as explicit as the ``yellow text box'' presentation shown within the \ConditionA{}, several participants saw this recommendation of ``unsupported'' factors as provocations themselves. It is possible that participants in \ConditionB{} benefited from these potentially provocative factors.


During pilot studies we observed that participants had difficulties interpreting items on the questionnaire measuring trust. To address this, we asked participants to think aloud during survey completion, and a researcher addressed queries and provided clarifications.

While our tasks were designed to be ecologically valid, as with any synthetic experiment, participants may not identify with the task domains, impacting their motivation to think critically. To mitigate this, our post-experiment interview asked participants to recall similar tasks from their own experience, when considering, e.g., how task importance might affect their critical thinking.

Some findings might be specific to our choice of shortlisting as a task paradigm and not representative of knowledge work tasks more broadly. Previous work has noted, e.g., a spectrum of approaches to integrating AI suggestions during creative writing \cite{singh2023hide}. Future work should explore alternative task paradigms, both in data-driven sensemaking (e.g., financial planning) and in other domains (e.g., design-based tasks).


Our interface has limitations. The table component is limited compared to commercial spreadsheets, for instance the inability to copy and paste blocks of data or to arbitrarily restructure the data, both important for sensemaking in spreadsheets \cite{joharizadeh2020gridlets,chalhoub2022freedom}. Rows could not be manually reordered, and we did not consider how manual and automatic ordering might be gracefully combined. This was ameliorated by allowing participants to add or remove arbitrary rows from their final shortlist. This did not inhibit our ability to study the effects of provocations on critical thinking, as the limitations were common to both experimental conditions and never explicitly cited by participants as being detrimental to the task.

Finally, the design of the provocations themselves are initial and provisional. We could have varied many design aspects, such as the length, timing, visual appearance, tone and grammatical style, positioning, and the content of provocations \cite{sarkar2024copilot}. As a starting point, we built on insights from previous work and employed vernacular idioms of comparable commercial GenAI interfaces such as Excel Copilot (Section~\ref{sec:SystemDesign}). Our experiment provides a simple comparison of presence vs. absence of provocations and provides evidence for the feasibility of the concept. Future work should compare different approaches to the presentation and content of provocations, with a view to improving their effectiveness at fostering critical thinking.

\section{Conclusion}
\label{sec:Conclusion}

This study explores the use of provocations -- AI-generated critiques and alternatives to AI suggestions -- for restoring critical thinking in AI-assisted knowledge work. Focusing on the task of shortlisting, we developed a system that generates factors for evaluating candidates and provides provocations to encourage critical reflection.

Our between-subjects study ($n=24$) revealed that provocations can have strong qualitative effects, inducing metacognition and critical thinking across all levels of Bloom's taxonomy. We derived five dimensions of the user experience of provocations: task urgency, task importance, user expertise, provocation actionability, and user responsibility.


While previous studies have explored critical thinking in education and specific professional domains, our work addresses the understudied area of common knowledge tasks. The findings highlight the potential of provocations to mitigate the risks of overreliance and mechanised convergence in AI-assisted work.

Future research could explore provocations for other knowledge workflows, investigate ways to quantitatively measure their impact, and refine the design of provocations based on the identified experiential dimensions. This work opens new avenues for designing AI systems that not only assist, but also enhance human critical thinking.


\bibliographystyle{ACM-Reference-Format}
\bibliography{references,references-eusprig,references-ppig}

\appendix
\input{appendices}

\end{document}

%% file: Tables/tbl_actioncodes.tex
\begin{table*}
\caption{Category codes for participant think-aloud utterances and actions.}
\label{tbl:actioncodes}
\begin{center}
\begin{tabular}{p{.10\linewidth} p{.75\linewidth}}
\toprule
{\small Code} & 
{\small Definition} \\
\midrule
{\small TA-FACT} & {\small Think-aloud -- reads text from or mentions factor card}\\
{\small TA-CRIT} & {\small Think-aloud -- reads text from or mentions factor criteria }\\
{\small TA-IMP} & {\small Think-aloud -- reads text from or mentions factor importance}\\
{\small TA-PROV} & {\small Think-aloud -- reads text from or mentions factor provocation (\ConditionA{} only)}\\
{\small TA-CONS} & {\small Think-aloud -- considers factor appropriateness or potential improvements}\\
{\small TA-GBSL} & {\small Think-aloud -- reads text from or mentions global shortlist}\\
{\small TA-FLSL} & {\small Think-aloud -- discusses how a specific factor affects shortlist}\\
{\small A-AFAC} & {\small Action -- Applies factor to shortlist}\\
{\small A-ECRIT} & {\small Action -- Edits factor criteria}\\
{\small A-EIMP} & {\small Action -- Edits factor importance}\\
{\small A-VGSL} & {\small Action -- Views global shortlist}\\
{\small A-ADDF} & {\small Action -- Adds new factor}\\
\bottomrule
\end{tabular}
\end{center}
\end{table*}

%% file: appendices.tex
\newpage
\section{Initial Factors and Provocations in User Study Tasks}
\label{Appendix:FactorsGenerated}
\input{Tables/tbl_factorsgenerated_CHARITIES}
\input{Tables/tbl_factorsgenerated_MOVIES}

\newpage
\section{Instruments}
\label{Appendix:Instruments}

\subsection{Reflective Thinking Questionnaire}
\label{Appendix:ReflectiveThinkingSurvey}

Adapted from \citet{kember2000development}. Responses are given on a 5-point Likert scale with the following options: A - definitely agree, B - agree with reservation, C - only to be used if a definite answer is not possible, D - disagree with reservation, and E - definitely disagree. Non-standard wording of the options comes from \citet{kember2000development}.

\begin{enumerate}
 \item When I am working on some activities, I can do them without thinking about what I am doing
 \item With this tool, the shortlist task requires me to understand the domain knowledge of the dataset used for shortlisting
 \item I sometimes question the way others do something and try to think of a better way
 \item As a result of using this tool, I have changed the way I look at myself
 \item When shortlisting with this tool, I did things without thinking about it,
 \item To make shortlists with this tool, you need to understand the field of knowledge related to the dataset
 \item I like to think over what I have been doing and consider alternative ways of doing it
 \item This tool has challenged some of my firmly held ideas
 \item As long as I can see the dataset and criteria for shortlisting in the tool, I do not have to think too much
 \item With this tool, I need to understand the dataset and the shortlist criteria in order to perform practical tasks
 \item I often reflect on my actions to see whether I could have improved on what I did
 \item As a result of using this tool, I have changed my normal way of doing things
 \item If I use this tool, I do not have to think too much about shortlisting
 \item With this tool, in the process of shortlisting, you have to continually think about the domain knowledge about the dataset
 \item I often re-appraise my experience so I can learn from it and improve for my next performance
 \item During the task, I discovered faults in what I had previously believed to be right
\end{enumerate}

\subsection{System Trust Questionnaire}
\label{Appendix:SystemTrustSurvey}

Used directly from \citet{jian2000foundations} without modification. Participants were given the following verbatim instructions: \emph{``Below is a list of statement for evaluating trust between people and automation. There are several scales for you to rate intensity of your feeling of trust, or your impression of the system while operating a machine. Please select the point which best describes your feeling or your impression. (Note: not at all=1: extremely=7)''}.

\begin{enumerate}
\item The system is deceptive
\item The system behaves in an underhanded manner
\item I am suspicious of the system's intent, action, or outputs
\item I am wary of the system
\item The system's actions will have a harmful or injurious outcome
\item I am confident in the system
\item The system provides security
\item The system has integrity
\item The system is dependable
\item The system is reliable
\item I can trust the system
\item I am familiar with the system
\end{enumerate}

\subsection{AI Workflow Reflective Thinking Questionnaire}
\label{Appendix:SystemImpactSurvey}

Responses are given on a 5-point Likert scale with the following options: strongly disagree, disagree, neither agree nor disagree, agree, strongly agree.

Your thinking during the task. Over the course of the task...
\begin{enumerate}
\item I updated my understanding of the domain 
\item I updated my understanding of my own preferences 
\item I updated my understanding of the AI system and its limitations 
\item I updated my understanding of the dataset and its limitations 
\item I learnt something about myself I didn't know before 
\item I learnt something about the domain I didn't know before 
\item I learnt something about the AI system that I didn't know before 
\item I checked the AI suggestions for errors 
\item I checked the AI suggestions for weaknesses or limitations 
\item I was critical or sceptical of the AI suggestions 
\item I considered the possibility that the AI suggestion could be wrong 
\item I considered the possibility that the AI suggestion could be improved 
\item The AI acted as an assistant, helping me achieve my task 
\item The AI acted as a critic, making me think about my task
\end{enumerate}

\subsection{Post-study Semi-Structured Interview Resources for \ConditionA{}}
\label{Appendix:InterviewQuestionBank}

Per standard practice of semi-structured interviews, this is not a definitive list that was asked exhaustively or sequentially, but served instead as a resource to help the researcher guide the discussion.

\begin{enumerate}
    \item What do you think of the factors listed in the prototype?  
    \item Did provocations help you think about which factors to consider? If yes, give an example and explain why. 
    \item Did provocations help you think about how to consider factors? If yes, give an example and explain why.
    \item Did provocations help make your decisions easier? Did they improve the final decision? 
    \item What kind of provocations do you think would be useful in helping you critically think about a data decision like the ones you saw today? 
    \item Are there other modalities of provocations that might be useful helping you critically think about your task? 
    \item Do you prefer provocations that are relevant to your dataset or ones that leverage outside data? 
    \item How important is actionability of provocations to you? 
    \item Where would you like to apply changes based on the provocations? To factors? To the original prompt?
    \item Was there anything surprising to you pertaining to the provocations? If yes, give an example and explain why.
    \item If you could change something about the interface, what would it be and why? 
    \item Can you think of an example in your own work or life to which you could apply this? 
    \item Do you see any challenges in applying a tool like this to your own work? 
\end{enumerate}

\newpage
\section{Prompts used in prototype}
\label{Appendix:Prompts}

\lstset{
    basicstyle=\footnotesize\ttfamily, 
    columns=fullflexible, 
    keepspaces=true, 
    lineskip=-0.5ex, 
}

\subsection{Generate factor prompt}
\label{Appendix:Prompts:GenerateFactors}

\begin{lstlisting}[basicstyle=\footnotesize]
"""\
I would like to select rows that fit these preferences: {shortlist_criteria}
I am currently considering these factors:
```json
{current_factors}
```

Please provide 3, 4 or 5 additional factors that I could consider to 
select rows that fit my preferences.
Factors must be unique. 
You may provide factors that are less relevant to my preferences 
but potentially inspiring and set their importance level to "Not at all".

Each factor must contain the following information:
- "name": The name of the factor or criteria. The name should be descriptive.
- "description": Describe how this factor can be directly available or 
inferred from existing columns. If not, describe how else people 
can collect this data about this factor.
- "available_in_data": Can I get the data directly from any existing columns? 
If so, true. If not, false.
- "source": Which columns can I directly analyze or I can extract data from?
- "suggestion": Suggest how to select rows based on this factor and why.
- "criteria": More concrete version of "suggestion" that is just about 
data threshold (e.g. for numbers, suggesting a concrete number threshold) 
or topics to extract (e.g. for text, suggesting general topics). 
No need to mention column names.
- "importance": How important this factor is; i.e. how much it should 
affect the final decision. 
Choose from the following option: "High", "Medium", "Low", "Not at all".
- "provocation": Highlight strengths, weaknesses, risks, limitations, 
alternatives, and biases. The risk of using such criteria, and what 
alternative criteria could be used. Suggest more relevant topics and 
keywords to the factor description. Even if there is no risk, suggest 
a case where the opposite of the criteria is better.

Your answer must be an array in JSON format, between triple backquotes. 
Here's the schema:
```
{factor_schema}
```\
"""

factor_schema = """\
{
    "name": str,
    "description": str,
    "available_in_data": bool,
    "source": list[str],
    "suggestion": str,
    "criteria": str,
    "importance": "High" | "Medium" | "Low" | "Not at all",
    "provocation": str
}\
"""
\end{lstlisting}

\subsection{Generate importance prompt}
\label{Appendix:Prompts:GetImportance}

\begin{lstlisting}[basicstyle=\footnotesize]
'''Task:
- Given lines of text and a description of topics or subjects, 
rate how relevant is each line to the provided topic.
- Extract all the lines that are relevant to the provided topic.
The text is given below as line number, text:
    {data_content}
- Give the output as a JSON list objects that match the following 
TypeScript interface:
```typescript
    interface Classification {{
    line: number;
    /**
    * These are EXACT quotes from the line 
    * that are related to the user description
    */
    key_terms: string[];
    relevance: "none" | "low" | "medium" | "high";
}}```
'''
\end{lstlisting}

\subsection{Apply factor prompt}
\label{Appendix:Prompts:ApplyFactors}

\begin{lstlisting}[basicstyle=\footnotesize]
default_get_analysis_content_prompt_template = """\
Please provide all the rows that fit my criteria, 
and reasons to include the rows.

My criteria: {factor_criteria}

Your answer must contain the following information:
- For each row, the row's "id_" and a "reason" to include the row.
- A "message" containing any warnings.

Your answer must be an object in JSON format, between triple backquotes. 
Here's the schema:
```
{response_schema}
```\
"""

analysis_response_schema = """\
{
    "data": list[{"id_": int, "reason": str}],
    "message": str
}\
"""
\end{lstlisting}

\subsection{Get overall recommendation (shortlist scoring) prompt}
\label{Appendix:Prompts:GetScores}
\begin{lstlisting}[basicstyle=\footnotesize]
"""\
I would like to select rows that fit these preferences: 
{shortlist_criteria}

I have already decided which factors to consider. 
Factors contain the following information:
- "factor_id": The factor's unique ID.
- "name": A short, descriptive name for the factor.
- "source": Which columns you can directly analyze or extract data from.
- "criteria": How to select rows based on this factor.

These are my factors:
```json
{current_factors}
```

Please decide which factor criteria each row meets. 
Your answer must contain the following information:
- For each row, the row's "id_" and the "factor_ids" 
of the factors whose criteria the row meets.
- A "message" containing any warnings.

Your answer must be an object in JSON format, between triple backquotes. 
It must include a record for every row. Here's the schema:
```
{match_schema}
```\
"""

match_schema = """\
{
    "data": list[{"id_": int, "factor_ids": list[str]}],
    "message": str
}\
"""
\end{lstlisting}

%% file: Tables/tbl_factorsgenerated_CHARITIES.tex
\begin{table*}[!hp]
\caption{Initial Factors Generated for the Charity Dataset}
\label{tbl:charityfactors}
\begin{center}

\begin{tabular}{p{.15\linewidth} p{.25\linewidth} p{.10\linewidth} p{.40\linewidth}}
{\footnotesize Name} & 
{\footnotesize Criteria} &
{\footnotesize Importance} &
{\footnotesize Provocation}\\
\midrule
{\footnotesize Target Population Mention} & {\footnotesize Keywords such as 'children', 'youth', 'families', 'young people'} & {\footnotesize High} & {\footnotesize May exclude grants that benefit children indirectly. Consider also looking for terms like 'education', 'school', 'playground', 'childcare'.}\\
{\footnotesize Amount of Grant} & {\footnotesize Higher numerical values} & {\footnotesize Medium} & {\footnotesize Larger grants may not always mean more impact on children. Consider the efficiency and scope of the project as well.}\\
{\footnotesize Specificity of Proposal} & {\footnotesize Detailed descriptions with specific activities or items.} & {\footnotesize Medium} & {\footnotesize Detailed descriptions may not always correlate with effectiveness. It's also subjective to determine what is considered 'detailed'.}\\
{\footnotesize Geographical Reach} & {\footnotesize Areas with schools, family communities, or known youth centers.} & {\footnotesize Low} & {\footnotesize Geographical data is not available in the dataset. External research is needed to determine areas with high child populations.}\\
{\footnotesize Past Performance} & {\footnotesize Reputation for successful child-focused projects.} & {\footnotesize Low} & {\footnotesize Data on past performance is not available in the dataset. May require significant additional research and could introduce bias towards established organizations.}\\
\bottomrule
\end{tabular}
\end{center}
\end{table*}

%% file: Tables/tbl_factorsgenerated_MOVIES.tex
\begin{table*}[!hp]
\caption{Initial Factors Generated for the Movies Dataset.}
\label{tbl:moviefactors}
\begin{center}

\begin{tabular}{p{.15\linewidth} p{.25\linewidth} p{.10\linewidth} p{.40\linewidth}}
{\footnotesize Name} & 
{\footnotesize Criteria} &
{\footnotesize Importance} &
{\footnotesize Provocation}\\
\midrule
{\footnotesize Family Genre} & {\footnotesize Include genres that are typically considered family-friendly.} & {\footnotesize High} & {\footnotesize Some movies may be categorized under family-friendly genres but still contain content that is not suitable for all family members. Always check the View Rating for additional guidance.}\\
{\footnotesize Low IMDb Score} & {\footnotesize IMDb Score below 5.0.} & {\footnotesize High} & {\footnotesize IMDb scores are subjective and may not always accurately reflect the quality of a movie. Some movies with low scores might have a cult following or be appreciated by specific audiences.}\\
{\footnotesize Low Rotten Tomatoes Score} & {\footnotesize Rotten Tomatoes Score below 40\%.} & {\footnotesize Medium} & {\footnotesize Rotten Tomatoes scores are based on critic reviews and may not align with audience opinions. Some movies may be critically panned but still enjoyable for certain viewers.}\\
{\footnotesize Child-Friendly Rating} & {\footnotesize View Rating is one of 'G', 'PG', 'TV-Y'.} & {\footnotesize High} & {\footnotesize Ratings are guidelines and may not always accurately represent the content's suitability for all children. Parental guidance is recommended.}\\
{\footnotesize Box Office Performance} & {\footnotesize Reputation for successful child-focused projects.} & {\footnotesize Low} & {\footnotesize Data on past performance is not available in the dataset. May require significant additional research and could introduce bias towards established organizations.}\\
\bottomrule
\end{tabular}
\end{center}
\end{table*}